\documentclass[11pt,a4paper]{article}

\usepackage[english,german]{babel}
\usepackage[latin1]{inputenc}
\usepackage[T1]{fontenc}
\usepackage{ae,aecompl}
\usepackage{times}
\usepackage{amsfonts,bbm,theorem}
\usepackage{amsmath}

\pdfoutput=0
\usepackage[dvips]{graphicx}

\usepackage{epsfig,url}

\newcommand{\myfigure}[2]{ \includegraphics*[#1]{#2} }

\newtheorem{theorem}{Theorem}[section]
\newtheorem{definition}[theorem]{Definition}
\newtheorem{corollary}[theorem]{Corollary}
\newtheorem{lemma}[theorem]{Lemma}
\newtheorem{proposition}[theorem]{Proposition}

\newenvironment{proof}{\begin{trivlist}\item[]{\em 
Proof:}\/}{\hfill\mbox{$\Box$}\end{trivlist}}
\newenvironment{proofof}[1]{\begin{trivlist}\item[]{\em Proof of #1:}\/}{%
\hfill\mbox{$\Box$}\end{trivlist}}

\newcommand{\Z}{{\mathbb Z}}
\newcommand{\R}{{\mathbb R}}
\newcommand{\C}{{\mathbb C\hspace{0.05 ex}}}
\newcommand{\vep}{\varepsilon}
\newcommand{\qand}{\quad\text{and}\quad}

\newcommand{\tf}{\tilde{h}}

\newcommand{\1}{\mathbbm{1}}
\newcommand{\rme}{{\rm e}}
\newcommand{\rmd}{{\rm d}}
\newcommand{\cou}{\beta}

\newcommand{\je}{j_{{\rm e}}}
\newcommand{\order}[1]{\mathcal{O}(#1)}
\newcommand{\braket}[2]{\langle #1 , #2\rangle}
\newcommand{\bigbraket}[2]{ \Big\langle #1 , #2\Big\rangle}
\newcommand{\mean}[1]{\langle #1\rangle}
\newcommand{\defem}[1]{{\em #1\/}} 
\newcommand{\defset}[2]{ \left\{ #1\left|\, #2
\makebox[0cm]{$\displaystyle\phantom{#1}$}\right.\!\right\} }
\newcommand{\set}[1]{\{#1\}}
\newcommand{\norm}[1]{\Vert #1\Vert}
\newcommand{\esssup}{\operatornamewithlimits{ess\,sup}}

\newcounter{jlisti}
\newenvironment{jlist}[1][(\thejlisti)]{\begin{list}{{\rm #1}\ \ }{ %
      \usecounter{jlisti} %
    \setlength{\itemsep}{0pt}
    \setlength{\parsep}{0pt}  %
    \setlength{\leftmargin}{0pt} %
    \setlength{\labelwidth}{0pt} %
    \setlength{\labelsep}{0pt} %
}}{\end{list}}

\DeclareMathOperator*{\sign}{sign}

\numberwithin{equation}{section}
\begin{document}
\selectlanguage{english}

\newcommand{\email}[1]{E-mail: \tt #1}
\newcommand{\emailjani}{\email{jlukkari@ma.tum.de}}
\newcommand{\addressjani}{\em Zentrum Mathematik,
Technische Universit\"at M\"unchen, \\
\em Boltzmannstr. 3, D-85747 Garching, Germany}

\newcommand{\emailherbert}{\email{spohn@ma.tum.de}}

\title{Anomalous energy transport in the FPU-$\beta$ chain} 
\author{Jani Lukkarinen\thanks{\emailjani},
  Herbert Spohn\thanks{\emailherbert}\\[1em]
\addressjani }

\maketitle

\begin{abstract}
We consider the energy current correlation function for the FPU-$\beta$
lattice.  For small non-linearity one can rely on kinetic theory.  The issue
reduces then to a spectral analysis of the linearized collision operator. We
prove thereby that, on the basis of kinetic theory, the energy current
correlations decay in time as $t^{-3/5}$. It follows that the thermal
conductivity is anomalous, increasing as $N^{2/5}$ with the system size $N$.
\end{abstract}


\section{Introduction and physical background}
\label{sec:intro}

With the availability of the first electronic computing machines, Fermi,
Pasta, and Ulam \cite{FPU55} investigated the dynamics of a chain of nonlinear
oscillators, in particular, their relaxation to thermal equilibrium.   Their
work had a, in retrospect surprisingly, strong impact.  We refer to the
special issue \cite{FPUfocus05} which accounts for the first fifty years.  In
our contribution, we will study the $\beta$-chain.   This is a linear chain of
equal mass particles which are coupled to their nearest neighbors by
nonlinear springs with a potential of the form 
$U_{\cou}(r) = \frac{1}{8} r^2 + \frac{1}{4} \cou r^4$, with $\cou>0$,
and $r$ being the string elongation.  
(According to the FPU convention, the $\alpha$-chain has a nonlinearity
$\frac{1}{3} \alpha r^3$ instead of $\frac{1}{4} \cou r^4$.)
If we denote the particles positions by
$q_i\in \R$, and their momenta (velocities) by $p_i\in \R$, then the
$\beta$-chain has the Hamiltonian 
\begin{align}
  H(q,p) = \sum_i \left[{\textstyle\frac{1}{2}} 
    p_i^2 + U_{\cou}(q_{i+1}-q_i)\right], 
\end{align}
and the dynamics is governed by 
\begin{align}\label{eq:Hamdyn}
  \frac{\rmd }{\rmd t} q_i = p_i, \quad
  \frac{\rmd }{\rmd t} p_i = U'_{\cou}(q_{i+1}-q_i) - 
  U'_{\cou}(q_i-q_{i-1}) .
\end{align}

Over the last decade there has been a lot of interest to understand the energy
transport through one-dimensional chains, amongst them the FPU $\beta$-chain
\cite{lepri01}. Numerically, one common setup is to consider a chain of length
$N$, and to couple its left- and rightmost particles to thermal reservoirs at
temperatures $T_-$ and $T_+$, respectively.   For long times the chain relaxes
to a steady state with a non-zero average energy current 
$\je (N)= (T_- - T_+) N^{-1}  \kappa(N)$, 
and the interest lies in the dependence of $\kappa(N)$ on $N$ for large $N$.
For a regular transport, i.e., for transport satisfying Fourier's law, one 
has $\kappa (N)\to {\rm const.}$ for large $N$.  Anomalous transport
corresponds to  
$\kappa (N)\simeq N^{\alpha}$, with $0<\alpha<1$.  In the $\beta$-chain
more recent
molecular dynamics simulations point to an $\alpha$ of approximately $0.4$
\cite{aoki01,LLP05}, and
thus a larger energy transport than expected on the basis of Fourier's law.  
In these simulations chain lengths of up to $N=2^{16}$ are used, and the
result seems to be stable for a range of fairly low boundary temperatures.
In \cite{MDN06} it is claimed that for somewhat higher boundary
temperatures, there is
a crossover at large $N$ to $\kappa (N)\simeq N^{\frac{1}{3}}$.  Hence, even
on the numerical level the accurate value of $\alpha$ is still being debated.

In this paper, we will adopt a different, but physically equivalent
procedure.   One prepares initially (for $t=0$) the infinite $\beta$-chain in
thermal equilibrium at temperature $T>0$.  This means that the initial
conditions of the Hamiltonian dynamics are distributed according to the 
(at this stage formal) Gibbs measure 
\begin{align}\label{eq:defGibbs}
Z^{-1} \rme^{-H/T} \prod_{i\in \Z} \left[ \rmd q_i  \rmd p_i\right].
\end{align}
This measure does not change in time.  One now adds some extra energy close to
the origin and studies the spreading of this excess energy.  To be more
precise, let us introduce the local energy, $e_i$, at the site
$i\in \Z$ by 
\begin{align}
 e_i(q,p) = \frac{1}{2} \left[ p_i^2 + U_{\cou}(q_{i+1}-q_i)
 + U_{\cou}(q_i-q_{i-1})\right].
\end{align}
We also employ the shorthand notation $e_i(t)=e_i(q(t),p(t))$, where
$(q(t),p(t))$ is the solution to the Hamiltonian dynamics (\ref{eq:Hamdyn})
for given initial conditions.  Then we define the normalized local average
excess energy by 
\begin{align}
 S(i,t) = \frac{1}{\chi} \left( \mean{e_i(t) e_0(0)}
 - \mean{e_i}\mean{e_0} \right).
\end{align}
Here $\mean{\cdot}$ denotes the thermal average (\ref{eq:defGibbs})
over the initial conditions, and $\chi = \sum_i 
\left( \mean{e_i e_0} - \mean{e_i}\mean{e_0} \right)$ is a
normalization guaranteeing $\sum_i S(i,0)=1$.  One has 
$S(-i,t)=S(i,t)$, and the
energy spread at time $t$ is defined as the spatial variance
\begin{align}
  D(t) = \sum_{i\in \Z} i^2 S(i,t).
\end{align}

Fourier's law corresponds to a diffusive spreading, $D(t) = \order{t}$ 
for large
$t$, while an exponent $\alpha>0$ corresponds to superdiffusive spreading
with $D(t) = \order{t^{1+\alpha}}$.  These properties can be more conveniently
reformulated by introducing for each directed bond from $i$ to $i+1$
a current $j_{i,i+1}$, 
so that the energy continuity equation holds in the following form:
\begin{align}
    \frac{\rmd }{\rmd t} e_i + j_{i,i+1} - j_{i-1,i}  = 0.
\end{align}
For the FPU-$\beta$ model such a current observable is given by
\begin{align}
 j_{i,i+1}(q,p) = -\frac{1}{2} (p_{i+1}+p_i) U'_{\cou}(q_{i+1}-q_i).
\end{align}
Obviously, $\mean{j_{i,i+1}}=0$.
We next introduce the energy current-current correlation function
\begin{align}
 C_{\cou}(t) = \sum_{i\in \Z} \mean{j_{0,1}(t) j _{i,i+1}(0)}.
\end{align}
Then
\begin{align}
  D(t) = D(0)+\frac{1}{\chi} \int_0^t \rmd s  \int_0^t \rmd s'
   C_{\cou}(s-s').
\end{align}
Note that $|C_{\cou}(t)| \le C_{\cou}(0)<\infty$.  Clearly, if 
$\int_0^\infty \rmd t  |C_{\cou}(t)|<\infty$, then $D(t)=\order{t}$.
On the other hand, if 
$C_{\cou}(t)= \order{t^{\alpha-1}}$ for large $t$ with $0<\alpha<1$,
then $D(t)=\order{t^{1+\alpha}}$, and the spreading is superdiffusive.

The problem of regular versus anomalous energy transport may thus be 
re\-phras\-ed
as whether $C_{\cou}(t)$ decays integrably or not.  Unfortunately, such a
reformulation is of little help.  To estimate the decay of a time correlation
in equilibrium, such as $C_{\cou}(t)$, is an exceedingly difficult problem.
However, in the limit of small $\cou$, through methods from kinetic
theory, $C_{\cou}(t)$ may be expressed in a more accessible form.  For the
complete argument we refer to \cite{Ziman67,ALS06,spohn06proc}. Here we only
state the small $\cou$ 
form of $C_\cou(t)$.  To do so will require some preparation.  But the
goal of our contribution is to  estimate the decay of  $C_{\cou}(t)$ for the
FPU-$\beta$ chain in the limit of small $\cou$.

At $\cou=0$, the system reduces to the harmonic Hamiltonian
\begin{align}
  H(q,p) = \sum_i \left[{\textstyle \frac{1}{2}} p_i^2 + 
    \textstyle{\frac{1}{8}} (q_{i+1}-q_i)^2\right],
\end{align}
which has the dispersion relation 
\begin{align}
\omega(k) = \sqrt{{\textstyle \frac{1}{2}}(1-\cos k)} =
\Bigl|\sin\frac{k}{2}\Bigr|.
\end{align}
Here we use the convention that the discrete Fourier transform yields
$2\pi$-periodic functions, and also declare that
the term ``periodic function'' always refers to a
function which is $2\pi$-periodic in all of its arguments.
It will be convenient to choose as the basic periodic cell the interval
$I=[0,2\pi)$.  In particular, then $x\bmod 2\pi \in I$ for all $x\in \R$.
On $I$ the dispersion relation is simply
\begin{align}
\omega(x)= \sin\frac{x}{2},
\end{align}
and thus also for all $x\ne 0$,
\begin{align}
\omega'(x) = \frac{1}{2} \cos\frac{x}{2},
\end{align}
and we let arbitrarily $\omega'(0)=0$.  We also introduce
\begin{align}
\Omega(x,y,z) = \omega(x)+\omega(y)-\omega(z)-\omega(x+y-z)
\end{align}
for $x,y,z\in \R$.  With these conventions the linearized collision operator
of the FPU-$\beta$ lattice in the kinetic limit is given by 
\begin{align}\label{eq:defL}
  (L f)(x) = \int_I\! \rmd y\int_I \!\rmd z\, \delta(\Omega(x,y,z))\,
  (f(x)+f(y)-f(z)-f(x+y-z)),
\end{align}
with $f$ periodically extended from $I$ to $\R$, see \cite{ALS06}. 

$L$ describes the collision
of two phonons, where $x,y$ label the incoming momenta and $z$, $x+y-z$ label
the outgoing momenta, thus by fiat satisfying momentum conservation
modulo $2\pi$.  Through the $\delta$-function the collisions are
also constrained to conserve energy.  Note that
at this stage, the definition in (\ref{eq:defL}) is
only formal since no prescription is given of how to deal with the
$\delta$-function.  It turns out to be useful to 
consider $\tilde{L}=\omega L\omega$ 
as a linear operator on $L^2(I)$, with 
$\omega$ being understood as the multiplication operator by the
function $\omega$.  We will prove later that $\tilde{L}$ 
is a bounded positive operator with a decomposition
\begin{align}
 \tilde{L}=W-A,
\end{align}
where $A$ is compact and $W$ is a multiplication operator.

Now we are in a position to state the conjectured behavior of
$C_{\cou}(t)$ for small coupling $\cou$.
\begin{trivlist}
\item
{\bf Kinetic conjecture:}\ \ {\it
For any $t\in \R$ and temperature $T>0$
\begin{align}
\lim_{\cou \to 0^+} C_{\cou}(\cou^{-2} t)
= \frac{T^2}{2\pi}
\bigbraket{\omega' }{
\exp\!\left[{-\pi^{-1} (12 T)^2 |t| \tilde{L}}\right] \omega' }, 
\end{align}
where $\braket{\cdot}{\cdot}$ denotes the scalar product in $L^2(I)$.
(A more detailed discussion about the scaling factors can be found in
\cite{ALS06}.) 
}
\end{trivlist}
Thus for small $\cou$, the decay of $ C_{\cou}(t)$ is obtained from the
spectral properties of $\tilde{L}$, certainly a more accessible item than the
full Hamiltonian dynamics.  Our goal here is to study the behavior of the
kinetic correlation function
\begin{align}
 C(t) =  \braket{\omega' }{\rme^{-|t| \tilde{L}} \omega' }.
\end{align}

In kinetic theory, it is a common practice to use the relaxation time
approximation, which in our case amounts to dropping the operator $A$, that
is, to approximate
\begin{align}
\braket{\omega' }{\smash{\rme^{-|t| \tilde{L}}} \omega' } \approx
\braket{\omega' }{\rme^{-|t| W} \omega' }.
\end{align}
As we will show, $W(x)=W(2\pi-x)$, and for $0<x\ll 1$, $W(x)$ behaves
asymptotically as $x^{5/3}$.  Thus the relaxation time approximation predicts
$\braket{\omega' }{\rme^{-|t| \tilde{L}} \omega' } = \order{t^{-3/5}}$ for
large $t$, as has been derived in \cite{perev03}.

$\tilde{L}$ has the range of $W$ as its essential spectrum.  In particular,
the essential spectrum starts from $0$.  Thus it is not obvious that the
asymptotics predicted by the relaxation time approximation is really the
correct one.  To understand the time decay
leads to two distinct  mathematical issues. 
\begin{jlist}
\item\label{it:collinvar} 
  The so called collisional invariants, which in essence are zero modes
  of $L$, could in principle prevent $C(t)$ from decaying to $0$. 
  To exclude such a possibility, we have to characterize all
  collisional invariants, which involves solving a non-trivial functional
  equation. 
\item We will use the resolvent expansion to estimate 
$\braket{\omega' }{\rme^{-|t| \tilde{L}} \omega' }$.  In our case, it turns out
to be necessary to make the expansion to the second order, yielding
\begin{align}\label{eq:resolv1a}
& \bigbraket{\omega' }{\frac{1}{\lambda +\tilde{L}} \omega' } 
= \bigbraket{\omega' }{\frac{1}{\lambda +W} \omega' }  + 
\bigbraket{\omega' }{\frac{1}{\lambda +W} A \frac{1}{\lambda +W} \omega' }  
\nonumber \\ & \qquad 
+ \bigbraket{\omega' }{\frac{1}{\lambda +W} A \frac{1}{\lambda +\tilde{L}} A
  \frac{1}{\lambda +W} \omega' } . 
\end{align}
The first term is identical to the relaxation time approximation, and behaves
as $\lambda^{-2/5}$ for $0<\lambda \ll 1$.  The second and third term will be
shown to be $\order{\lambda^{-1/5-\vep}}$ for any $\vep>0$.  Although also this
second contribution is divergent, the first term is dominant, and thus we
confirm the prediction of the relaxation time approximation in this particular
case.
\end{jlist}

An inherent difficulty in resolvent expansions is the estimation of the
remainder term, such as the last term in (\ref{eq:resolv1a}).
Our method bears some similarity to the Birman-Schwinger 
estimates used in quantum mechanics.  It relies on the fact that 
the resolvent expansion is made up to an even order, 
as well as on the operator $B=W^{-1/2} A W^{-1/2}$ being compact.  In fact,
it is likely that similar techniques can be used to study many of the 
cases where a decomposition $\tilde{L}=W-A$, with $W\ge 0$ and a compact $B$,
is possible, although we would expect the optimal order for the resolvent
expansion to vary from case to case. 
The exact order, as well as the exact power
of the decay, would naturally depend also on the function $\omega'$.
A reader interested in such generalizations is invited to jump ahead to
the proof of the main theorem in Section \ref{sec:resolv}.

Our results imply that, on the kinetic time scale, the energy spread is
superdiffusive, with $D(t) \simeq c\, t^{7/5}$, $c>0$, for large $t$.
This corresponds to a heat conduction exponent $\alpha=\frac{2}{5}$ and is in
agreement with the molecular dynamics simulations of \cite{aoki01,LLP05}.
As the example of long time tails in classical fluids teaches us, kinetic
theory might miss the true asymptotic decay of equilibrium correlation
functions.  Whether this is the case also for the FPU-$\beta$ chain, remains
a challenge for the future.

From the point of view of kinetic theory, our result is fairly
surprising. Usually linearized collision operators have a spectral gap
implying exponential decay of the current-current correlation function, and
diffusive spreading for the corresponding conserved quantity.  In fact, we are
not aware of any other Boltzmann type kinetic model which would exhibit
superdiffusive spreading.

\subsection*{Acknowledgements}

We would like to thank Walter Aschbacher, 
Laurent Desvillettes, Antti Kupiainen, Michael Loss, Cl\'{e}ment Mouhot,
Andrey Pereverzev, and Gennadi Vainikko for useful comments and references.
Many of these exchanges happened during a stay at the
Erwin Schr\"{o}din\-ger International Institute for Mathematical Physics (ESI),
Vienna, Austria, whom we thank for their hospitality.  
This work has been completed as part of the
Deutsche Forschungsgemeinschaft (DFG) project SP~181/19-2.

\section{Main results}
\label{sec:mainres}

To define $L$, we first need to find all solutions to the energy constraint.
The solution manifold to $\Omega(x,y,z)=0$, is clearly 
non-empty, as there are the \defem{trivial solutions}
\begin{align}
  z=x \qand z=y.
\end{align}
We will later prove in Corollary \ref{th:ysols} that, in addition, there is
a solution $y=h(x,z)$, and that all
other solutions are modulo $2\pi$ equal to one of these three.
For $x,z\in I$ the function $h$ is given by
\begin{align}\label{eq:hfinalsol}
  h(x,z) = \frac{z-x}{2} +
  2 \arcsin \! \left( \tan \frac{|z-x|}{4} \cos \frac{x+z}{4} \right)
\end{align}
where  $\arcsin$ denotes the principal branch with values in 
$[-\pi/2,\pi/2]$.  We extend $h$ to $\R^2$ by defining 
\begin{align}\label{eq:hsolcont}
  h(x,z) = h(x \bmod 2\pi,z \bmod 2\pi) -i(x) ,
\end{align}
where $i(x)= x-(x \bmod 2\pi)\in 2\pi\Z$.
This choice makes $h$ everywhere continuous while ensuring that
for all $x,z\in \R$, we still  have $\Omega(x,h(x,z),z)=0$.

The energy conservation $\delta$-function can then
be formally resolved by integrating over some chosen direction:
for instance, choosing the $y$-integral for this purpose
would yield for any $z\ne x$ and for any continuous periodic function $G$,
\begin{align}
& \int_I\! \rmd y\, \delta(\Omega(x,y,z)) G(x,y,z) 
\nonumber \\ & \quad 
= \frac{1}{|\partial_2\Omega(x,z,z)|} G(x,z,z)
+ \frac{1}{|\partial_2\Omega(x,h(x,z),z)|} G(x,h(x,z),z).
\end{align}
However, this procedure is somewhat suspect here, as it will lead to terms of
the type $\infty-\infty$, related to the trivial solutions and canceled only
due to symmetry properties.  An additional difficulty lies in the
application of the definition to functions $G$ which are not continuous but
merely $L^2$-integrable.  To put the definition of $L$ on a firmer ground, we 
will resort to a different approach in Section \ref{sec:FPUopdef}: 
we replace $\delta$ in (\ref{eq:defL})
by a regularized $\delta$-function
$\delta_\epsilon(X) = \epsilon \pi^{-1} (\epsilon^2 + X^2)^{-1}$,
$\epsilon>0$, and then
show that there is a unique self-adjoint operator $L$ which agrees 
with these operators in the limit $\epsilon\to 0$.  Our choice of
regularization for the $\delta$-function is not completely arbitrary:  
in the kinetic limit of lattice systems with random mass perturbations 
the corresponding $\delta$-function also appears via a sequence
of $\delta_\epsilon$-functions  (see, for instance, Proposition A.1 in
\cite{ls05}). 

A somewhat lengthy computation, to be discussed in Sections
\ref{sec:FPUopdef} and \ref{sec:FPUoperator}, shows that the
formal procedure explained before
is essentially correct: the trivial solutions give no contribution, 
and the unique limit operator $L$ is
\begin{align}\label{eq:defL1}
 L = V+K_1 - 2 K_2,
\end{align}
where $K_1$ and $K_2$ are integral operators determined by the
integral kernels
\begin{align}\label{eq:defKs}
  K_1(x,y) & =
  4\frac{\1(F_-(x,y)>0)}{\sqrt{F_-(x,y)}} \qand
  K_2(x,y) =  \frac{2}{\sqrt{F_+(x,y)}} ,
\end{align}
which are defined for $x,y\in I$ using the auxiliary functions
\begin{align}\label{eq:defFpm}
F_{\pm}(x,y) = \left(\cos \frac{x}{2} + \cos \frac{y}{2}\right)^2 \pm 
    4 \sin \frac{x}{2} \sin \frac{y}{2}.
\end{align}
In addition, $V$ denotes a multiplication operator by the function
\begin{align}\label{eq:defV}
 V(x) = \int_I \rmd y\, K_{2}(x,y).
\end{align}
$L$ was already used as the linearized collision operator
in \cite{perev03}.
In addition to $L$, $\tilde{L} = \omega L \omega$, and $W=\omega^2 V$,
the operator $B=W^{-1/2} (W-\tilde{L}) W^{-1/2}$ will be of importance.
Explicitly, $B$ is then defined via the integral kernel 
\begin{align}\label{eq:defkerB}
B(x,y) = V(x)^{-1/2} (2 K_2(x,y)-K_1(x,y)) V(y)^{-1/2}.
\end{align}

Let us next list the main properties of these operators, to be proven
in Sections \ref{sec:FPUoperator} and \ref{sec:provecolli}.  
We start with the results related to
item \ref{it:collinvar} mentioned in the introduction.
\begin{definition}
A measurable periodic function
$\psi:\R\to \C$ is called a \defem{collisional invariant} 
if for almost every $x,y,z\in \R$ such that $\Omega(x,y,z)=0$,
\begin{align}\label{eq:defcollinv}
\psi(x)+\psi(y)-\psi(z)-\psi(x+y-z) =0.
\end{align}
\end{definition}
In the definition, ``almost every'' refers to the Lebesgue measure on
any two-dimensional submanifold of the full
solution set.
The following theorem shows that, in the case considered here, there are 
only the obvious collisional invariants.
\begin{theorem}\label{th:collinv} 
Suppose $\psi$ is periodic and locally integrable:
$\psi|_I \in L^1(I)$. Then $\psi$ 
is a collisional invariant if and only if there are 
$c_1,c_2\in \C$ such that $\psi(x)= c_1+ c_2 \omega(x)$  for 
a.e.\  $x$.
\end{theorem}
In higher dimensions there is a general argument which identifies the
collisional invariants under minimal assumptions on $\omega$ \cite{spohn06}.
In contrast, our proof here relies heavily on the specific form of $\omega$,
and does not exclude the appearance of non-trivial collisional invariants 
in some other
one-dimensional systems.

\begin{definition}
We define a \defem{parity transformation} $P:L^2(I)\to L^2(I)$
by letting $(P\psi)(0)=\psi(0)$ and, for $x\in (0,2\pi)$,
\begin{align}
 (P\psi)(x) = \psi(2 \pi -x).
\end{align}
\end{definition}
Clearly, $\omega(x)$ is symmetric, and $\omega'(x)$ is antisymmetric under $P$.

\begin{proposition}\label{th:operprop}
$\tilde{L}$ is a bounded positive operator, and
$B$ is a compact self-adjoint operator on $L^2(I)$.  
Both $B$ and $\tilde{L}$ commute with the parity operator $P$.
In addition, $B\le 1$, and
$B\psi =\psi$ if and only if the periodic extension of 
$V^{-1/2}\psi$ is a collisional invariant. 
\end{proposition}

\begin{theorem}\label{th:mainres}
Let $R:(0,\infty)\to \R_+$ be defined by
\begin{align}
R(\lambda) = \bigbraket{\omega' }{\frac{1}{\lambda+\tilde{L}} \omega' }.
\end{align}
Then there is $0<c_0<\infty$ such that with $\alpha =\frac{2}{5}$
\begin{align}\label{eq:Rlimit}
\lim_{\lambda\to 0^+} \lambda^{\alpha } R(\lambda) = c_0.
\end{align}
\end{theorem}
Since $R(\lambda)=\int_0^\infty \rmd t\, \rme^{-\lambda t} C(t)$,
for $\lambda>0$, $R(\lambda)$ is a Laplace transform of the monotonically
decreasing positive function $C(t)$.  Methods from Tauberian theory 
can then be used to connect the asymptotic behavior of $R$ and $C$,
proving that the asymptotic decay of the
current-current correlations is given by $C(t) = \order{t^{-\frac{3}{5}}}$,
and that the integrated correlations grow like
$ \int_0^{t}\! \rmd s\, C(s) = \order{t^{\frac{2}{5}}}$.
Explicitly,
\begin{corollary}\label{th:maincorr}
With $c_0>0$ and $\alpha =\frac{2}{5}$ as in Theorem \ref{th:mainres},
and with $\Gamma$ denoting the gamma function, we have
  \begin{align}\label{eq:Ctdecay}
    \lim_{t\to \infty} t^{1-\alpha } C(t) = \frac{c_0}{\Gamma(\alpha )}.
  \end{align}
\end{corollary}
(For a proof of the result, see 
for instance ``Zusatz zu Satz 2'' on p.\ 208 of \cite{doetsch37}.)

We have divided the proof of the above results in four sections.
We solve the energy constraint and derive the above form for the operator $L$
in Section \ref{sec:FPUopdef}.  Proposition \ref{th:operprop}
is proven in Section \ref{sec:FPUoperator}, which includes,
in particular, the estimates proving the compactness of $B$.  
We study the collisional invariants in Section \ref{sec:provecolli}, and 
prove Theorem  \ref{th:collinv}  there.  
Finally, these results are then applied in a resolvent
expansion, and we prove Theorem \ref{th:mainres} in Section \ref{sec:resolv}.
The short Appendix contains a convenient estimate for the norm of an integral
operator.

\section{Resolution of the energy constraint}
\label{sec:FPUopdef}

We will define the operator $L$ by the following procedure:
we consider a regularization of the $\delta$-function by
\begin{align}
\delta_\epsilon(X) = \frac{\epsilon}{\pi} \frac{1}{\epsilon^2 + X^2},
\end{align}
for any $0<\epsilon\le 1$.  Let $L_\epsilon$ denote the operator defined by
(\ref{eq:defL})  
after $\delta$ has been replaced by $\delta_\epsilon$.  This yields a
bounded operator, for which using the symmetry properties of the integrand
\begin{align}\label{eq:Lbqf}
& 4 \braket{f}{L_\epsilon f} = 
 \int_{I^3} \rmd x \rmd y \rmd z\, \delta_\epsilon(\Omega(x,y,z))
  |f(x)+f(y)-f(z)-f(x+y-z)|^2.
\end{align}
Our aim in this section is to prove the following result about the limiting
behavior of this quadratic form when $\epsilon\to 0^+$.
\begin{proposition}
\label{th:Lbetaprop}
For any $f:\R\to \C$, which is periodic and Lipschitz continuous,
the limit $\lim_{\epsilon\to 0^+} \braket{f}{L_\epsilon f}$ exists, and it is
non-negative, finite, and equal to 
\begin{align}\label{eq:Ltlowerb}
& \int_{I^2} \rmd x \rmd z \,\frac{1}{2 \sqrt{F_+(x,z)}}
  \left|f(x)+f(h(x,z))-f(z)-f(x-z+h(x,z))\right|^2 
 \nonumber \\ & \quad
=  \int_{I^2} \rmd x \rmd z \,f(x)^* (V(x)+K_1(x,z)-2 K_2(x,z))f(z).
\end{align}
\end{proposition}
The proof of the Proposition will require some fairly technical estimates not
needed later,  
and a reader accepting our definition of the operator $L$ and the equality in
(\ref{eq:Ltlowerb}) can well
skip the proofs of the Lemmas below in the first reading.

We will begin by constructing the solutions to the energy constraint 
$\Omega=0$, and then study the behavior of $\Omega$ around this set, in order
to evaluate the limit of the approximate $\delta$-functions.
Let $D=[0,2\pi]^3$ be the closure of $I^3$.  For $(x,y,z)\in D$,
\begin{align}
\Omega(x,y,z) = \sin\frac{x}{2} + \sin\frac{y}{2} -
\sin\frac{z}{2} - \left|\sin\frac{x+y-z}{2}\right|.
\end{align}
Since then $-\pi\le \frac{x+y-z}{2}\le 2\pi$, we can split $D$
into two sets
$U_+$ and $U_-$, depending on the sign of the last term. Explicitly, 
\begin{align}
 U_+ & = \defset{(x,y,z)\in D}{x+y-2\pi\le z\le x+y},\\
 U_- & = \defset{(x,y,z)\in D}{x+y\le z\text{ or }z\le x+y-2\pi},
\end{align}
and $\Omega(x,y,z) =\Omega_\sigma(x,y,z)$ with
$\sigma = +1$ if $(x,y,z)\in U_+$, and 
with $\sigma=-1$ if $(x,y,z)\in U_-$, where
\begin{align}
\Omega_\sigma(x,y,z) =
 \sin\frac{x}{2} + \sin\frac{y}{2} -
\sin\frac{z}{2} - \sigma \sin\frac{x+y-z}{2}.
\end{align}
The following representations of these functions will become useful later
(they can be checked, for instance, by expressing the
trigonometric functions in terms of complex exponentials):
for all $x,y,z\in\R$,
\begin{align}
 \Omega_+(x,y,z) & = 4 \sin \frac{x-z}{4}\sin \frac{y-z}{4}\sin \frac{x+y}{4},
\label{eq:Ompluseq} \\ 
\Omega_-(x,y,z) & = 2 \left( 
   \cos \frac{x+z}{4} \sin \frac{x-z}{4} +
   \cos \frac{x-z}{4} \sin \frac{2 y + x-z}{4} \right).
\label{eq:Omminuseq}
\end{align}
From these, we directly find the zeroes of $\Omega$:
\begin{lemma}\label{th:Omzerolemma}
Let $Z = \defset{(x,y,z)\in D}{\Omega(x,y,z)=0}$, 
and denote $Z_\pm = Z\cap U_\pm$.  $Z_+$ consists of those 
$(x,y,z)\in D$  for which either $z=x$, $z=y$, or $y=x \in \set{0,2\pi}$.
$Z_-$ consists of those $(x,y,z)\in D$ which satisfy any of the following
three conditions, where $h$ is defined by (\ref{eq:hfinalsol}) 
and (\ref{eq:hsolcont}),
\begin{enumerate}
\item $x=0$, $z=2\pi$, or $x=2\pi$, $z=0$,
\item $x \le z$, and $y=h(x,z)$,
\item $x \ge z$, and $y=2\pi + h(x,z)$.
\end{enumerate}
In addition, 
for $(x,y,z)\in U_-$, with $x\ne z$, we have
$\sign(z-x)\partial_y \Omega_-(x,y,z)\ge 
\cos^2 \frac{x-z}{4}$.
\end{lemma}
\begin{proof}
By (\ref{eq:Ompluseq}), $\Omega=0$ on $U_+$ if and only if one of the three
factors is zero.  Since $|\frac{a-b}{4}|\le \frac{\pi}{2}$,
and $0\le \frac{a+b}{4}\le \pi$,
for any $a,b\in\set{x,y,z}$, this can be checked to coincide with the above 
classification of $Z_+$.

To compute $Z_-$, let us first consider the case $\cos \frac{x-z}{4}=0$.  Then
either $x=0$, $z=2\pi$, or $z=0$, $x=2\pi$, and both cases
can be checked to form solutions for any $y$.  Otherwise,
$\cos \frac{x-z}{4}>0$.  Also
$\cos\frac{x-z}{4} \ge \cos \frac{x+z}{4}$, as 
$|\frac{x-z}{4}|\le \frac{x+z}{4}\le \pi$.
Similarly, as $|\frac{x-z}{4}|\le \frac{2\pi-x+2\pi-z}{4}\le \pi$,
we have $\cos\frac{x-z}{4} \ge -\cos \frac{x+z}{4}$.  Therefore, 
$|\cos \frac{x+z}{4}| \le\cos\frac{x-z}{4}$.  Also by (\ref{eq:Omminuseq}) 
\begin{align}\label{eq:ommprime}
 \partial_y \Omega_-(x,y,z) = 
   \cos \frac{x-z}{4} \cos \frac{2 y + x-z}{4}.
\end{align}  
We then split the proof into three steps with additional conditions on $x,z$.

Assume first $z= x$.  Then $(x,y,z)\in U_-$ if and only if
$y=0$ or $y=2\pi$, and both cases clearly yield solutions.
Since $h(x,x)=0$, both cases are covered by the Lemma.

Assume then $z> x$.  Then $(x,y,z)\in U_-$ if and only if
$0\le y\le z-x<2\pi$.  This implies that 
$|\frac{2 y + x-z}{4}|\le \frac{z-x}{4} < \frac{\pi}{2}$,
and thus in this case 
$\partial_y \Omega_- \ge \cos^2 \frac{x-z}{4} >0$.  On the other hand,
by explicit computation, then 
$\Omega_-(x,0,z)\le 0$ and $\Omega_-(x,z-x,z)\ge 0$.  Therefore, for such
$x,z$ there is a \defem{unique} solution $y\in [0,z-x]$.  
By (\ref{eq:Omminuseq}) this solution satisfies
\begin{align}\label{eq:2yxz}
\sin \frac{2 y + x-z}{4} =
   \cos \frac{x+z}{4} \tan \frac{z-x}{4}.
\end{align}
This equation has infinitely many solutions $y\in\R$, but
the above bounds show that exactly one of them,
\begin{align}
y & =  \frac{z-x}{2} + 2 \arcsin \Bigl[
   \cos \frac{x+z}{4} \tan \frac{z-x}{4} \Bigr] = h(x,z),
\end{align}
with $\arcsin$ denoting the principal branch with values in 
$[-\pi/2,\pi/2]$, can belong to $[0,z-x]$.  Since there must be a solution in
this interval, we find that $h(x,z)\in [0,z-x]$, and thus
$(x,h(x,z),z)\in U_-$.

To complete the analysis, assume $z< x$.  Then 
$(x,y,z)\in U_-$ if and only if $2\pi+z-x\le y\le 2\pi$. 
We let $x'=2\pi -x$, etc., when $z'>x'$, and $0\le y'\le z'-x'$.
As  always $\Omega_-(x',y',z') = \Omega_-(x,y,z)$, we can conclude that
for any $x,z$ there is a unique solution in $U_-$ which satisfies
$y'=h(x',z')$, i.e., the solution is
\begin{align}
y & = 2\pi + \frac{z-x}{2} + 2 \arcsin \Bigl[
   \cos \frac{x+z}{4} \tan \frac{x-z}{4} \Bigr] = 2\pi + h(x,z).
\end{align}
It also follows that in this case, 
$\partial_y \Omega(x,y,z)\le -\cos^2 \frac{x-z}{4}<0$
for all $2\pi+z-x\le y\le 2\pi$.  This completes the proof of the Lemma.
\end{proof}
As $\Omega$ is periodic, 
the Lemma yields immediately also a classification of the zeroes of
$\Omega$ in $\R^3$.
\begin{corollary}\label{th:ysols}
$\Omega(x,y,z)=0$ if and only if at least one of the following equalities
holds modulo $2\pi$: $z=x$, $z=y$, or $y=h(x,z)$.
\end{corollary}
\begin{proof}
It is clear from the Lemma that any $(x,y,z)\in \R^3$ satisfying the above
condition is a zero of $\Omega$.  For the converse,
assume $\Omega(x,y,z)=0$.  Then for $x'= x\bmod 2\pi$, etc., also
$\Omega(x',y',z')=0$, and we can apply the Lemma.  If
$(x',y',z')\in U_+$, 
then either $z'=x'$, $z'=y'$ or $x'=y'=0$.  Since the last condition
implies $z'=0$, and thus $y'=0=h(0,0)$, also the last instance is covered in
the Corollary.
If the point belongs to $U_-$, we must have $y'=h(x',z')\bmod 2\pi$, 
and thus then $y=h(x,z)$ modulo $2\pi$.
\end{proof}

The following Lemma can then be used to compute the relevant limits 
for integrals over $U_-$:
\begin{lemma}\label{th:UminusGlim}
Suppose $G:\R^3\to \C$ is a periodic continuous function.
Then
\begin{align}\label{eq:Flim}
& \lim_{\epsilon\to 0^+} \int_{U_-} \rmd x \rmd y \rmd z 
  \, \delta_\epsilon(\Omega(x,y,z)) G(x,y,z)
 \nonumber \\ & \quad
=  \int_{I^2} \rmd x \rmd z \frac{2}{\sqrt{F_+(x,z)}}
  G(x,h(x,z),z).
\end{align}
\end{lemma}
\begin{proof}
As $G$ is periodic and continuous, it is also bounded.  Let $0<\vep<1$ be
arbitrary, and denote $X_\vep = [0,\vep]\times [2\pi-\vep,2\pi] \cup 
[2\pi-\vep,2\pi]\times [0,\vep]$.  Let us first consider
some fixed $x,z\in [0,2\pi]^2\setminus X_\vep$, $x\ne z$.  
By Lemma \ref{th:Omzerolemma},
$|\partial_y \Omega|\ge \cos^2\frac{x-z}{4}\ge \sin^2\frac{\vep}{4}>0$,
and $y\mapsto \Omega(x,y,z)$ is a bijection with a unique zero, $y_0$, which is
equal to $h(x,z)$ modulo $2\pi$.  We change the integration variable $y$
to $t=\Omega(x,y,z)/\epsilon$, 
which shows that the integral over $y$ is equal to
\begin{align}
  \int_{a/\epsilon}^{b/\epsilon} \rmd t\, \frac{1}{|\partial_2 \Omega(x,y(\epsilon t),z)|}
 \frac{1}{\pi(1+t^2)}  G(x,y(\epsilon t),z).
\end{align}
with $a\le 0$ and $b\ge 0$ and $y(0)=y_0$.  This is always bounded by 
a constant which depends on $\vep$ but not on $\epsilon$.
Thus an application of the dominated convergence theorem shows that
\begin{align}\label{eq:Flima}
& 
\lim_{\epsilon\to 0^+} \int_{U_-} \rmd x \rmd y \rmd z \1((x,z)\not\in X_\vep)
  \, \delta_\epsilon(\Omega(x,y,z)) G(x,y,z)
 \nonumber \\ & \quad
=  \int_{I^2} \rmd x \rmd z 
\frac{\1((x,z)\not\in X_\vep)}{|\partial_2 \Omega_-(x,h(x,z),z)|}
  G(x,h(x,z),z).
\end{align}
We used here the observation 
that the set of $x= z$, as well as that of $(x,z)$ for which $a=0$ or
$b=0$, have zero measure.
Here, by (\ref{eq:ommprime}), we have 
\begin{align}\label{eq:OmmmmFplus}
& |\partial_2 \Omega_-(x,h(x,z),z)| = 
\left|\cos \frac{x-z}{4}\right| \left(
1- \sin^2 \frac{2 h + x-z}{4}\right)^{\frac{1}{2}}
 \nonumber \\ & \quad
= \left(\cos^2 \frac{x-z}{4}- 
\sin^2 \frac{x-z}{4} \cos^2 \frac{x+z}{4}\right)^{\frac{1}{2}}
= \frac{1}{2} \sqrt{F_+(x,z)},
\end{align}
where the last equality can be checked by a calculation, for instance,
using the identity
$\cos^2 u = \frac{1}{2} (1+\cos(2 u))$.  
For all $x,z\in I$, we clearly have an estimate
\begin{align}\label{eq:K2bound}
0\le K_2(x,z) =  \frac{2}{\sqrt{F_+(x,z)}} \le
\left(\sin\frac{x}{2}\sin\frac{z}{2}\right)^{-1/2}.
\end{align}
Thus $F_+^{-1/2}$ is integrable, and 
we can again apply dominated convergence to take the limit 
$\vep\to 0$ inside the integral.  This proves that the right
hand side of (\ref{eq:Flima}) converges to the right hand side of 
(\ref{eq:Flim}).

Therefore, to complete the proof of the Lemma, it is sufficient to prove
that the integral over $(x,z)\in X_\vep$ vanishes when first $\epsilon\to 0$
and then $\vep \to 0$.  In fact, using the symmetry between the two
components of $U_-$ and the boundedness of $G$, it is 
sufficient to study the integral
\begin{align}
J_\vep = 
\int_0^\vep \rmd x \int_{2\pi-\vep}^{2\pi} \rmd z
\int_0^{z-x} \rmd y\, \frac{\epsilon}{\pi} \frac{1}{\epsilon^2+\Omega_-^2}.
\end{align}
We split the integral over $y$ into two parts at $y=\pi$.
If $0\le y\le \pi$, we have $2 \partial_z \Omega_- \ge -\cos\frac{z}{2}
\ge \cos\frac{\vep}{2}$.  Therefore, we can perform the $z$ integral first,
as above, and conclude that the result of the $z$-integral is uniformly
bounded in $\epsilon$ and $\vep$.  
Performing then the $x$ and $y$ integrals, shows that
the full integral is bounded by a constant times $\vep$.
In the remaining region, $\pi\le y\le z-x$, we have
$2 \partial_x \Omega_- \ge \cos\frac{x}{2}\ge \cos\frac{\vep}{2}$.
Thus in this case, we can perform the $x$ integral first, with a  uniformly
bounded result.  Performing then the $y$ and $z$ integrals, and combining the
bound with the earlier estimate, proves that
there is $c>0$ such that $J_\vep \le c \vep$.  Thus 
$J_\vep \to 0$ as $\vep \to 0$, which concludes the proof of the Lemma.
\end{proof}

To complete the proof of Proposition \ref{th:Lbetaprop}, we require 
one more Lemma, closely related to the above estimates. 
\begin{lemma}\label{th:lemmaK2toK1}
Assume $G$ is measurable and periodic on $\R^2$.  Then
\begin{align}\label{eq:lemmaK2toK1}
\int_{I^2}\rmd x\rmd z\, \frac{1}{\sqrt{F_+(x,z)}} G(x,h(x,z))
= \int_{I^2}\rmd x\rmd y\, 2 \frac{\1(F_-(x,y)>0)}{\sqrt{F_-(x,y)}} G(x,y),
\end{align}
as long as either $G\ge 0$, or one of the above integrals is absolutely
convergent. 
\end{lemma}
\begin{proof}
As is apparent from (\ref{eq:lemmaK2toK1}),
the proof is accomplished by a change of integration variables
from $z$ to $y=h(x,z)$ for a fixed $x$.  However, even computing the local 
inverse functions from (\ref{eq:hfinalsol}) does not appear to be completely
straightforward.  We will resort to a roundabout way, which relies on the
fact that $h(x,z)$ is a solution to the energy constraint on $\Omega_-$.

If $\tf(y;x)$ is a local inverse of $h(x,\cdot)$, then for all $y$ in its
domain there is $n\in \set{0,1}$ such that $(x,y+2\pi n,\tf(y;x))\in U_-$ and
\begin{align}
\Omega(x,y,\tf(y;x))=\Omega_-(x,h(x,\tf(y;x)),\tf(y;x))=0.
\end{align}
Conversely, assume that $x\in (0,2\pi)$ is given, 
and $\tf(y;x)$ is a map from some interval $J\subset(0,2\pi)$ to $I$
such that either $x+y\le \tf(y;x) \le 2\pi$ or
$0\le \tf(y;x) \le x+y-2\pi$ for all $y$, and $\Omega(x,y,\tf(y;x))=0$.
Then by Lemma \ref{th:Omzerolemma}, we must have in the first case
$y=h(x,\tf(y;x))$, and in the second case, $y=2\pi+h(x,\tf(y;x))$
for all $y$.  

Therefore, to construct all possible local inverse functions
of $h(x,\cdot)$, it is sufficient to find for given $x,y$ all $z$
such that $(x,y,z)\in U_-$, and $\Omega_-(x,y,z)=0$.
We begin from the following representation of $\Omega_-$: for all $x,y,z$,
\begin{align}
\Omega_-(x,y,z) & = 2 \left( 
   \sin \frac{x+y}{4} \cos \frac{x-y}{4} +
   \cos \frac{x+y}{4} \sin \frac{x+y-2 z}{4} \right).
\end{align}
Let us assume that $(x,y,z)\in U_-$ with $0<x<2\pi$.  
Then $\cos \frac{x+y}{4}=0$ implies $y=2\pi-x$, and thus then
$\Omega_-= 2\sin \frac{x}{2} >0$.  
Thus if $\Omega_-=0$, we have
\begin{align}\label{eq:sineq2}
\sin \frac{x+y-2 z}{4}= - \tan \frac{x+y}{4} \cos \frac{x-y}{4} .
\end{align}
Since $x,y,z$ are real, this is possible only if the absolute value of 
the right hand side is less than or equal to
one.  This condition is equivalent to the condition $F(x,y)\ge 0$,
with
\begin{align}
F(x,y) = \cos^2 \frac{x+y}{4}- \sin^2 \frac{x+y}{4} \cos^2 \frac{x-y}{4}.
\end{align}
A brief computation reveals
that, in fact, $F(x,y)= \frac{1}{4} F_-(x,y)$, and thus $F_-\ge 0$ is a 
necessary condition to have any solutions.

When $F_-(x,y)\ge 0$, (\ref{eq:sineq2}) holds if and only if
there is $n\in \Z$ such that $z=z_n$, where
\begin{align}
z_n = 2\pi n + \frac{x+y}{2} + (-1)^{n} 2 \arcsin 
\left(\tan \frac{x+y}{4} \cos \frac{x-y}{4} \right),
\end{align}
with $\arcsin$ denoting the principal branch.  There can be maximally two
values of $n$ for which $z_n$ belongs to $[0,2\pi)$.  However, by inspecting
the sign of $\partial_3\Omega_-$, 
similarly to what was done in the proof of Lemma \ref{th:Omzerolemma},
we find that for $F_->0$ there are exactly two solutions in $U_-$, and that
for a given $x$
there are maximally two values of $y$ for which $F_-=0$.  If $F_->0$, the
solutions are explicitly $z=\tilde{h}_\pm(y;x)$, 
where for either choice of the sign $\sigma \in\set{\pm 1}$
\begin{align}
 \tilde{h}_\sigma(y;x) = \frac{x+y}{2} + c_\sigma 2\pi +
\sigma 2 \arcsin 
\left(\tan \frac{x+y}{4} \cos \frac{x-y}{4} \right),
\end{align}
and $c_\sigma=0$, if $\sigma=+1$, and $c_\sigma=(-1)^{\1(x+y>2\pi)}$, if
$\sigma=-1$.   

Therefore,
apart from a finite number of values $y\in[0,2\pi]$, there are either no, or
there are exactly two,
solutions in $U_-$.  Both of the 
solutions satisfy (\ref{eq:sineq2}) and thus also for any such $z$
\begin{align}
&|\partial_3 \Omega(x,y,z)| = 
   \left|\cos \frac{x+y}{4}\right| 
   \left(1-\sin^2 \frac{x+y-2 z}{4} \right)^{\frac{1}{2}}
\nonumber \\ & \quad 
 = \sqrt{F(x,y)} = \frac{1}{2} \sqrt{F_-(x,y)}.
\end{align}
On the other hand, by implicit differentiation we find
\begin{align}
\partial_2 \Omega_-(x,h(x,z),z) \partial_z h(x,z)
+ \partial_3 \Omega_-(x,h(x,z),z)  = 0.
\end{align}
By (\ref{eq:OmmmmFplus}) this implies
\begin{align}
\frac{1}{2} \sqrt{F_+(x,z)}\left| \partial_z h(x,z) \right| 
= \left|\partial_3 \Omega_-(x,h(x,z),z)  \right| ,
\end{align}
which allows to compute the Jacobian of the change of variables.

Collecting all of the above results together, and applying
Fubini's theorem, we can conclude that (\ref{eq:lemmaK2toK1}) holds
for $G\ge 0$ and for any $G$ which is bounded.  This implies, in particular,
that the integrals are equal if $G$ is replaced by $|G|$ for any measurable
$G$.  Thus if either of these integrals in (\ref{eq:lemmaK2toK1}) is
absolutely  convergent, then the other must be
so as well.  Then an application of dominated convergence theorem proves that 
(\ref{eq:lemmaK2toK1}) holds also for such measurable $G$.
\end{proof}

\begin{proofof}{Proposition \ref{th:Lbetaprop}}
Let $f$ be a periodic Lipschitz function.  We express
$\braket{f}{L_\epsilon f}$ as an integral over $I^3$ using (\ref{eq:Lbqf}).
As $D =U_+\cup U_-$, and 
$D\setminus I^3$ and $U_+\cap U_-$ have measure zero,
we can split the integral into two parts by using 
$\int_{I^3} =\int_{U_+} + \int_{U_-}$.
Since the factor multiplying $\delta_\epsilon$ in the integrand is positive,
periodic, and continuous, Lemma \ref{th:UminusGlim} implies that the integral
over $U_-$ converges to the left hand side of (\ref{eq:Ltlowerb}).
By boundedness of $f$ and applying (\ref{eq:K2bound}), 
the integral yields a finite, non-negative result as claimed in the
Proposition. 

Thus in order to prove convergence to the left hand side of
(\ref{eq:Ltlowerb}), we only need to show that the integral over $U_+$
vanishes as $\epsilon\to 0^+$.
For any $(x,y,z)\in U_+$, using (\ref{eq:Ompluseq}) and the fact that
$|\sin x|\ge \frac{2}{\pi}|x|$ for $|x| \le \frac{\pi}{2}$,
\begin{align}
& |\Omega_+(x,y,z)| \ge 4 \frac{|x-z|}{2 \pi}\frac{|y-z|}{2 \pi}
 \sin\frac{x+y}{4} 
\ge   \frac{1}{2 \pi^{3}} |x-z| |y-z| m(x,y),
\end{align}
where $m(x,y) = \min(x+y,4\pi-x-y)$.
This implies that, if $|x-z|\le |y-z|$, then by the Lipschitz property of
$f$, there is a constant $c>0$ such that
\begin{align}
& \frac{1}{\epsilon^2+\Omega_+^2} 
  |f(x)+f(y)-f(z)-f(x+y-z)|^2 
\le 
 \frac{c}{m(x,y)^2} \frac{|x-z|^2}{(\pi \epsilon)^2+|x-z|^4 } .
\end{align}
If $|y-z|\le |x-z|$, the same estimate holds after $x$ and $y$ have been
interchanged on the right hand side.  Therefore,
\begin{align}\label{eq:Upest}
& \int_{U_+} \rmd x \rmd y \rmd z\, 
 \frac{\epsilon}{\pi} \frac{1}{\epsilon^2 + \Omega(x,y,z)^2}
  |f(x)+f(y)-f(z)-f(x+y-z)|^2
\nonumber \\ & \quad
\le  c \int_{I^2} \rmd x \rmd y  \frac{\1(m(x,y)\ge \epsilon)}{m(x,y)^2}
 \int_{-\infty}^\infty \rmd t\, 
  \frac{\epsilon}{\pi} \frac{t^2}{(\pi \epsilon)^2+t^4 } 
 + c' \epsilon^2 \epsilon^{1-2},
\end{align}
where the second term is an estimate for the integral over $(x,y)$ with
$m(x,y)<\epsilon$ --- these are contained in the two boxes $[0,\epsilon]^2$
and $[2\pi-\epsilon,2\pi]^2$ and we have there estimated the integrand
trivially using $\Omega^2\ge 0$.
The first integral over $x,y$ is $\order{|\ln \epsilon|}$, and the second
integral is 
$\order{\epsilon^{1/2}}$, as seen by changing the integration variable to
$s=\epsilon^{-1/2} t$.  Therefore, we can conclude that the left hand side of
(\ref{eq:Upest}) vanishes as $\epsilon\to 0^+$, i.e., that the integral over
$U_+$ does not contribute to limit, as long as $f$ is
a Lipschitz function.  

Thus to complete the proof of the Proposition, we only need to prove the
equality in (\ref{eq:Ltlowerb}).  Instead of doing this directly, let us come
back to the integral over $U_-$, which was 
proven above to converge to the left hand side of (\ref{eq:Ltlowerb}).  
The set $U_-$ is clearly invariant under $x\leftrightarrow y$.  
By inspection we check that this is also true for the 
map $z\mapsto z'$, with $z'=x+y-z+2\pi$, for $x+y\le z$, and
$z'=x+y-z-2\pi$, otherwise.  Similarly, $U_-$ is left invariant
under the map $(x,y,z)\mapsto (x',y',z')$, with 
$y'=z$, $z'=y$, and $x'=x+y-z+2\pi$, for $x+y\le z$, and
$x'=x+y-z-2\pi$, otherwise.  All of these maps leave also $\Omega$ invariant.
We can thus first expand the square and then use the above mappings to 
appropriately change variables to prove that
\begin{align}
& \frac{1}{4} \int_{U_-} \rmd x \rmd y \rmd z\, 
\delta_\epsilon(\Omega(x,y,z))
  |f(x)+f(y)-f(z)-f(x+y-z)|^2
\nonumber \\ & \quad
= \int_{U_-} \rmd x \rmd y \rmd z\, 
\delta_\epsilon(\Omega(x,y,z))
 f(x)^* (f(x)+f(y)-2 f(z)).
\end{align}
Lemma \ref{th:UminusGlim} can be applied to the right hand side proving
that it converges to 
\begin{align}
& \int_{I^2} \rmd x \rmd z\, 
 \frac{2}{\sqrt{F_+(x,z)}} f(x)^* (f(x)+f(h(x,z))-2 f(z))
\end{align}
By Fubini's theorem, the first and the last terms are equal to those of the
right hand side of (\ref{eq:Ltlowerb}).  The middle term is absolutely
convergent by (\ref{eq:K2bound}), and thus an application of Lemma
\ref{th:lemmaK2toK1} shows that it 
is equal to the missing $K_1$-term in (\ref{eq:Ltlowerb}).  
This completes the proof of the Proposition. 
\end{proofof}

\section{Linearized collision operator}
\label{sec:FPUoperator}

We will derive in this section the regularity properties of $\tilde{L}$ and
$B$ and prove Proposition \ref{th:operprop}.
Let us recall the definition of $L_\epsilon$ in Section \ref{sec:FPUopdef},
and Proposition \ref{th:Lbetaprop} proven there.  Let  also
$\tilde{L}_\epsilon = \omega L_\epsilon \omega$.  If $f$ is a periodic
Lipschitz function, then so is $g=\omega f$, and thus an immediate consequence
of the Proposition is that then
\begin{align}\label{eq:Ltildelim}
\lim_{\epsilon\to 0^+}\braket{f}{\smash{\tilde{L}_\epsilon} f} =
\lim_{\epsilon\to 0^+}\braket{g}{L_\epsilon g} =
\braket{f}{\smash{\tilde{L}} f} \ge 0.
\end{align}
Here we have employed the definition of $\tilde{L}$ to identify it in the
right hand side of (\ref{eq:Ltlowerb}).  We shall soon prove that $\tilde{L}$
is a bounded operator on $L^2(I)$.  As Lipschitz functions are dense in
$L^2(I)$, the above result implies that $\tilde{L}$ is a positive operator.
Moreover, $\tilde{L}$ is then \defem{uniquely} determined
by (\ref{eq:Ltildelim}) in the following sense:  Suppose $L'$ is another
self-adjoint operator (not necessarily bounded) for which 
$\braket{f}{L' f}=\braket{f}{\smash{\tilde{L}}f}$ for every Lipschitz
function $f$.   Since then $L'-\tilde{L}$ is densely defined with
$\braket{f}{(L'-\smash{\tilde{L}})f}=0$, we can conclude using the polarization
identity that $L'f=\tilde{L}f$ for all $f$ Lipschitz.  As $\tilde{L}$
is \defem{bounded} and self-adjoint, this implies $L'=\tilde{L}$.
Thus we only need to check that $\tilde{L}$ is bounded, and to show that
$[P,\tilde{L}]=0$, in order 
to conclude the properties stated about   $\tilde{L}$
in Proposition \ref{th:operprop}.  
We remark in passing that only $\tilde{L}$ will be proven to be bounded, 
the operator $L$ could well be unbounded.

For the proof of compactness of $B$, we need more precise estimates on $W$ and
on the kernel functions $K_1$ and $K_2$.  We recall the definition of $W$
given in Section \ref{sec:mainres}, $W = \omega^2 V$.
As to be shown,  the exponent $\alpha=\frac{2}{5}$ 
in the main theorem
is determined by the behavior of $W(x)$ near $x=0$.  This will be
summarized in the following Lemma.
\begin{lemma}\label{th:Wtest}
The function $W:\R \to \R_+$ is symmetric, $P W=W$, and
continuous. In addition, there are
constants $c_1, c_2>0$, such that for all $x\in \R$
\begin{align}\label{eq:Wtbounds}  
  c_1 \left|\sin \frac{x}{2}\right|^{\frac{5}{3}} \le W(x)
\le  c_2 \left|\sin \frac{x}{2}\right|^{\frac{5}{3}},
\end{align}
and also 
$\lim_{x\to 0} \left(|\sin \frac{x}{2}|^{-5/3} W(x)\right) = w_0 \in (0,\infty)$,
where
\begin{align}
w_0 = 4 \int_{0}^{\infty}\!\rmd s\, \left(2 s + s^4\right)^{-\frac{1}{2}}.
\end{align}
\end{lemma}
\begin{proof}
If $x'=2\pi-x$, then $\cos\frac{x'}{2}=-\cos\frac{x}{2}$,
and $\sin \frac{x'}{2}=\sin\frac{x}{2}$.  
Therefore, 
$F_\pm(2\pi-x,2\pi-y)=F_\pm(x,y)$, and a change of variables shows that
$(PW)(x)=W(x)$ for all $x$.  
We will soon prove that the function 
\begin{align}
f(x)= \omega(x)^{-\frac{5}{3}} W(x) =\omega(x)^{\frac{1}{3}}
\int_0^{2\pi} \rmd y\, \frac{2}{\sqrt{F_+(x,y)}},
\end{align}
is continuous, with $f(0)=w_0>0$.  This implies that $f$ has a minimum and
maximum on $[0,2\pi]$, and since $f(x)>0$, the minimum is non-zero.
This will directly imply that $W$ is
continuous and satisfies the bounds in (\ref{eq:Wtbounds}).  
Therefore, to complete the proof of the Lemma, we only need to study $f$.

Suppose $x\in (0,2\pi)$.  Then the bound (\ref{eq:K2bound}) allows
using the dominated convergence theorem to prove that 
$\lim_{h\to 0} f(x+h)=f(x)$, which proves that $f$ is continuous at $x$.
Thus we only need to prove that $f$ is continuous at $x=0$ and $x=2\pi$
and, as $f(2\pi-x)=f(x)$ for all $x$, it suffices to study the limit 
$x\searrow 0$.  Assume thus $0<x<\pi/4$.  Then for all 
$0\le y \le \frac{3}{2}\pi$, 
we have $\cos(x/2)+\cos (y/2) \ge \cos(\pi/8)-\cos(\pi/4)>0$, and thus also
$F_+(x,y)\ge C>0$.  Therefore, 
\begin{align}
\lim_{x\to 0^+} f(x) = \lim_{x\to 0^+}
\int_0^{\frac{\pi}{2}} \rmd y\,
\frac{2 s_x^{\frac{1}{3}}}{\sqrt{(c_x-c_y)^2+4 s_x s_y}},
\end{align}
where $s_x = \sin(x/2)$, $c_x=\cos(x/2)$, etc.
In the final integral, let us denote $\vep = s_x$, and 
change variables to $s= \vep^{-1/3} \sin\frac{y}{2}$.  This shows that
the integral is equal to 
\begin{align}
\int_0^{\vep^{-\frac{1}{3}} 2^{-\frac{1}{2}}} \rmd s\, 
\frac{4 \vep^{\frac{1}{3}+\frac{1}{3}}}{\sqrt{1-\vep^{2/3} s^2}} 
\left[4 \vep^{1+\frac{1}{3}} s + \Bigl(\frac{\vep^{2/3} s^2 -\vep^2}{
    \sqrt{1-\vep^{2/3} s^2}+\sqrt{1-\vep^2}}\Bigr)^2
 \right]^{-\frac{1}{2}}.
\end{align}
Here $\vep^{2/3}$ can be canceled between the two
factors.  Now $x\to 0^+$ implies $\vep\to 0^+$, and the limit 
can also be taken directly 
from the integrand, as a straightforward application of 
the dominated convergence theorem will show.  Therefore, we can conclude 
(with a final change of variables to $s/2$) that 
\begin{align}
\lim_{x\to 0^+} f(x) =
\int_0^{\infty} \rmd s\, \frac{4}{\sqrt{4 s + \frac{1}{4} s^4}} = w_0.
\end{align}
Clearly, $w_0$ is strictly positive and finite, and thus by defining 
$f(2\pi n)=w_0$,  
$f$ becomes a function which is continuous everywhere.  This completes the
proof of the Lemma.
\end{proof}

\begin{figure}
  \begin{center}
    \myfigure{width=0.6\textwidth}{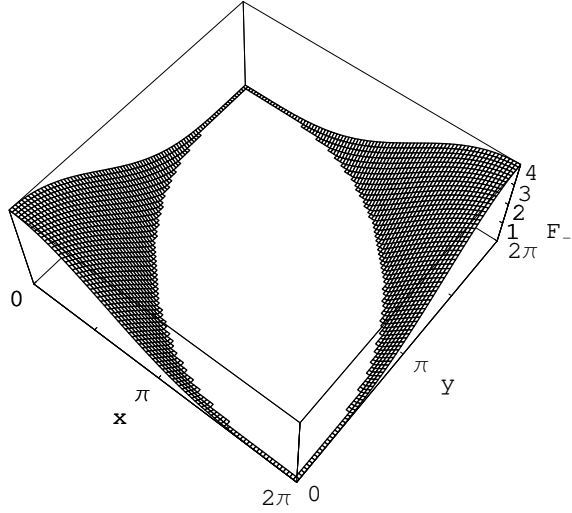}
    \caption{Plot of the positive part of $F_-(x,y)$.}
    \label{fig:Fminus}
  \end{center}
\end{figure}

We then require information about structure of singularities of the kernels
$K_1$ and $K_2$ defined in (\ref{eq:defKs}).  $K_2(x,z)$ is bounded apart from
the point singularities at $(x,z)=(0,2\pi)$ and $(2\pi,0)$, 
and estimate (\ref{eq:K2bound}) will suffice to control its behavior.
In contrast, $K_1(x,y)$ has two line singularities of strength
$\frac{1}{2}$, which coalesce at the corners $(x,y)=(0,2\pi)$ and
$(2\pi,0)$ forming a point singularity of strength $1$.  To control these
singularities, we will resort to the 
estimates given in the following Lemma.
For the sake of illustration, we have plotted the positive part of  
$F_-$ in Fig.\ \ref{fig:Fminus}.  
\begin{lemma}
\label{th:mainFpmest}
Let $x\in (0,2\pi)$ be given. Then there are $y_1,y_2$ such that
$0<y_1<2\pi-x<y_2<2\pi$, $F_-(x,y)\le 0$ for $y_1\le y\le y_2$, and
\begin{align}
F_-(x,y)&\ge C (y_1-y)\sin\frac{x}{2} ,\quad \text{for}\quad 0\le y < y_1,
\label{eq:Fmsmall}\\
F_-(x,y)&\ge C (y-y_2)\sin\frac{x}{2} ,\quad \text{for}\quad y_2<y\le 2\pi,
\label{eq:Fmbig}
\end{align}
with a constant $C>0$ independent of $x,y$.
\end{lemma} 
\begin{proof}
As $F_-(2\pi-x,2\pi-y)= F_-(x,y)$, it suffices to prove the Lemma
for $0< x \le \pi$.  For notational simplicity, let
$c=\cos \frac{x}{2}$ and $s=\sin \frac{x}{2}$. Then 
$0<s\le 1$ and $c=\sqrt{1-s^2}\in [0,1)$.

Since $F_-$ is continuous and
$F_-(x,0)=(1+c^2)^2>0$, $F_-(x,2\pi)=(1-c^2)^2>0$, and 
$F_-(x,2 \pi-x)=-s^2<0$, we can find $0<y_1<2\pi-x<y_2<2\pi$
such that $F_-(x,y_i)=0$ and $F_-(x,y)\le 0$ for $y_1\le y\le y_2$.
Assume then that $0\le y < y_1$, when $0<x+y<2\pi$.
As $F_-(x,y_1)=0$, we have
\begin{align}
F_-(x,y)=-\int_{y}^{y_1} \rmd z\, \partial_2 F(x,z),
\end{align}
and to complete the proof of (\ref{eq:Fmsmall}), it will be sufficient to show
that $\partial_2 F_-(x,y) \le -C s$ for all $0\le y < y_1$.
Similarly, to prove (\ref{eq:Fmbig}), it suffices to show that 
$\partial_2 F_-(x,y) \ge C s$ for all $y_2<y\le 2\pi$.

\begin{figure}
  \begin{center}
    \myfigure{width=0.6\textwidth}{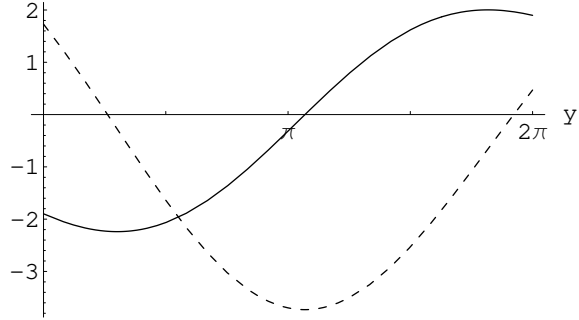}
    \caption{Plot of $\partial_2 F_-(x,y)$ (solid line) and of 
      $F_-(x,y)$ (dashed line) for $x=2.5$.}
    \label{fig:F2plot}
  \end{center}
\end{figure}

Let us thus consider the function
\begin{align}
F_2(y) = \partial_2 F_-(x,y) =
 -\sin\frac{y}{2} 
(c + \cos \frac{y}{2}) - 2 s \cos \frac{y}{2}.
\end{align}
We claim that there are $y'_\pm$ such that $F_2(y)$
is strictly decreasing for $0\le y\le y'_-$
and for $y'_+\le y\le 2\pi$, and 
strictly increasing for $y'_-\le y\le y'_+$.  As $F_2(0)=-2s<0$
and $F_2(2\pi)=2s>0$, then there is a unique $y'_1$ 
such that $F_2(y'_1)=0$.  Then $y'_-<y'_1<y'_+$,
$F_2(y)<0$ for $y<y'_1$, and 
$F_2(y)>0$ for $y>y'_1$.  Since $F_-$ is then strictly decreasing up to 
$y=y'_1$ and after that strictly increasing, we can conclude that
$y_1<y'_1<y_2$, and that $F_-(x,y)>0$ for $y<y_1$ and for $y>y_2$.
Therefore, to complete the proof of
the Lemma, we only need to find $0<C\le 2$ such that
$\partial_2 F_-(x,y_1) \le -C s$, and 
$\partial_2 F_-(x,y_2) \ge C s$.  To make the above argument more transparent,
we have plotted a sample $F_2$ in Fig.\ \ref{fig:F2plot}.

Let us first consider estimating 
$\partial_2 F_-(x,y_2)$. Since $y_2 > 2\pi -x$, now
$\cos\frac{y_2}{2}< -c$, and thus
\begin{align}\label{eq:y2eq}
 \cos\frac{y_2}{2}+c = -2\sqrt{s \sin\frac{y_2}{2}}.
\end{align}
Therefore, denoting $t_2=\sin \frac{y_2}{2}$,
\begin{align}
 F_2(y_2) = 2 \sqrt{s} t_2^{3/2} +2 s \sqrt{1-t_2^2}.
\end{align}
Thus if $t_2^2\le \frac{1}{2}$, $F_2(y_2) \ge \sqrt{2} s$.
But also when $t_2^2\ge \frac{1}{2}$,
$F_2(y_2) \ge 2^{1-3/4} \sqrt{s}\ge 2^{1/4} s$.  Therefore, always
$\partial_2 F_-(x,y_2)\ge 2^{1/4} s$.

We estimate $\partial_2 F_-(x,y_1)$ next.
Since $y_1<2\pi -x$, we have 
$\cos\frac{y_1}{2}> -c$, and
\begin{align}\label{eq:y1eq}
 \cos\frac{y_1}{2}+c = 2\sqrt{s \sin\frac{y_1}{2}}.
\end{align}
Let $\vep>0$ be sufficiently small.
If $y_1\le \pi-2 \vep$, then
$\cos\frac{y_1}{2}\ge \sin\vep>0$, and thus $F_2(y_1)\le -2 s\sin\vep$.
If $|y_1-\pi|\le 2 \vep$, then $|\cos\frac{y_1}{2}|\le \sin\vep$,
and $\sin\frac{y_1}{2}\ge \cos\vep$.  Thus, by (\ref{eq:y1eq}),
\begin{align}
F_2(y_1)\le -2 s \left[(\cos\vep)^{\frac{3}{2}}-\sin\vep\right].
\end{align}
Therefore, choosing $\vep$ sufficiently small 
(for instance, $\vep=\frac{1}{2}$)
we have again obtained a bound of the required type.

We have thus proved the result for $y_1\le \pi+2 \vep$.  Assume then
$y_1>\pi+2 \vep$, and let $t_1=\sin \frac{y_1}{2}$.  Then
$\cos \frac{y_1}{2} = -\sqrt{1-t_1^2}$, and (\ref{eq:y1eq}) implies that
\begin{align}
2 \sqrt{s t_1} = c- \sqrt{1-t_1^2} = \frac{t_1^2-s^2}{c+\sqrt{1-t_1^2}}
\le \frac{t_1^2}{2\sqrt{1-t_1^2}}.
\end{align}
Therefore, $t_1^{3/2}\ge 4 \sqrt{1-t_1^2} \sqrt{s}$, and thus
\begin{align}
& F_2(y_1) = -2 s^{\frac{1}{2}} t_1^{\frac{3}{2}}+ 2 s \sqrt{1-t_1^2} 
 \le -2 s \sqrt{1-t_1^2} (4-1) \le -6 s \sin \vep .
\end{align}
This proves that there is a pure constant $C>0$ such that 
$\partial_2 F_-(x,y_1) \le -C s$.

We still need to prove the monotonicity property of $F_2$ mentioned earlier.
Let $u=\cos\frac{y}{2}$, when $u$ goes from $1$ to $-1$ strictly monotonicly,
as $y$ goes from $0$ to $2\pi$.  Also
\begin{align}
  F_2(y) = - 2 s u - (c+u) \sqrt{1-u^2} =g(u).
\end{align}
Then
\begin{align}
 g'(u) = -2 s - \sqrt{1-u^2} + \frac{u (c+u)}{\sqrt{1-u^2}}, \quad
 g''(u) = \frac{c+3 u-2 u^3}{(1-u^2)^{3/2}}.
\end{align}
The polynomial $c+3 u-2 u^3$ has a local minimum at $u=-2^{-1/2}$ and 
a local maximum at $u=2^{-1/2}$.  Its values at $u=-1,0,1$ are
$c-1,c,c+1$, respectively.  Thus there is $-1<u_0\le 0$ such that
$g''(u_0)=0$, $g''(u)<0$ for $u<u_0$ and $g''(u)>0$
for $u>u_0$.  Since $c<1$, $g'(u)$ first 
decreases strictly from $+\infty$ to $g'(u_0)$ and then increases strictly to 
$+\infty$ again.  Since $g'(0)<0$, also $g'(u_0)<0$ and thus there are 
$u_\pm$ such that $g$ is strictly increasing for $-1\le u\le u_-$
and for $u_+\le u\le 1$ and strictly decreasing for $u_-\le u \le u_+$.
This implies the stated monotonicity property of $F_2$ and completes the
proof of the Lemma.
\end{proof}

We then prove two
intermediate compactness results which will become useful in the proof of
Proposition \ref{th:operprop}.
Instead of Sobolev-space techniques (such as proving that the operators improve
a Sobolev index), we will rely on direct norm estimates
which are quite straightforward in the present case.
\begin{proposition}\label{th:K2compact}
Let $\psi:I\to \C$ be given, and assume that there are $C,p>0$ such that
\begin{align}\label{eq:psibound}
|\psi(x)|\le C \left(\sin\frac{x}{2}\right)^p 
\end{align}
for all $x\in I$. Then the function 
\begin{align}
  K(x,y) = \psi(x)^* K_2(x,y)  \psi(y)
\end{align}
defines a compact, self-adjoint integral operator on $L^2(I)$.
\end{proposition} 
\begin{proof}
Since $\left(\sin\frac{x}{2}\right)^{p-1/2}\in L^2(I)$, the estimate
(\ref{eq:K2bound}) proves that 
$K$ is Hilbert-Schmidt, and thus also compact.
As $K(x,y)$ is symmetric, the operator is self-adjoint.
\end{proof}

\begin{proposition}\label{th:K1compact}
Let $\psi:I\to \C$ satisfy the assumptions of Proposition \ref{th:K2compact}.
Then the function 
\begin{align}
  K(x,y) = \psi(x)^* K_1(x,y)  \psi(y)
\end{align}
defines a compact, self-adjoint integral operator on $L^2(I)$.
\end{proposition}
\begin{proof}
Let $p,C>0$ be constants for which (\ref{eq:psibound}) holds.
As the bound is a decreasing function of $p$, it is sufficient to prove the 
Proposition assuming $0<p\le \frac{1}{2}$.
Suppose $0<\vep<2\pi$ is arbitrary, and let $T_\vep$ denote the integral
operator defined by 
\begin{align}
&  X_\vep = \defset{(x,y)\in [\vep,2\pi-\vep]^2}{
  F_-(x,y)\ge C\vep\sin\frac{\vep}{2}},\\
 & T_\vep(x,y) = \1((x,y)\in X_\vep) K(x,y).
\end{align}
Then $|T_\vep(x,y)|\le c(\vep\sin\frac{\vep}{2})^{-1/2}$ for some constant
$c$, and since the kernel
$T_\vep(x,y)$ is obviously symmetric, we can conclude that $T_\vep$ is a
self-adjoint Hilbert-Schmidt operator on $L^2(I)$.
As $K(x,y)$ is also symmetric, we can apply Proposition \ref{th:Anormcor}
to the integral operator $K-T_\vep$. We will choose
$\phi(x)=\sin(x/2)^p$, $\alpha = 1-\frac{1}{2p}$, and
$A(x,y) = \sin(x/2)^{-p} (K(x,y)-T_\vep(x,y))\sin(y/2)^{-p}$. 
By (\ref{eq:psibound}), then
$|A(x,y)| \le C^2 \1((x,y)\not\in X_\vep) K_1(x,y)$.
Therefore, it is enough to inspect the integral
\begin{align}
J(x) = \int_{I}\! \rmd y\,
\left(\sin\frac{x}{2}\right)^{p+\frac{1}{2}} 
\left(\sin\frac{y}{2}\right)^{p-\frac{1}{2}} 
\1((x,y)\not\in X_\vep) K_1(x,y).
\end{align}
We claim that there is $c>0$, such that
$J(x)\le c \vep^{p}$ for all $x\in I$.  
Then by Proposition \ref{th:Anormcor}, 
we have $\norm{K-T_\vep}\le C^2 c \vep^{p}$.
Since this also holds for $\vep>\pi$, when $T_\vep=0$, we find that $K$ is
itself a self-adjoint bounded operator.  However, then also
$T_\vep\to K$ in norm, and since each $T_\vep$ is a compact
operator, and the space of compact operators is closed in the operator norm,
we conclude that $K$ is also compact, proving the results mentioned in the
theorem.  

Thus we only need to show that $J(x)\le c \vep^p$ for all $x\in I$, assuming
$0<p\le \frac{1}{2}$.
For any $x\in I$ we obtain from Lemma \ref{th:mainFpmest}
the following rough estimate, where the integration region 
is estimated trivially,
\begin{align}
 & J(x) \le C^{-1/2} \left(\sin\frac{x}{2}\right)^{p} \Bigl(
\int_0^{y_1} \rmd y\, f_1(y) 
+ \int_{y_2}^{2\pi} \rmd y\, f_2(y)\Bigr) ,
\end{align}
where
\begin{align}
f_1(y) & =
\left(\sin\frac{y}{2}\right)^{p-\frac{1}{2}} \frac{1}{\sqrt{y_1-y}}\qand
f_2(y) =\left(\sin\frac{y}{2}\right)^{p-\frac{1}{2}} \frac{1}{\sqrt{y-y_2}}.
\end{align}
Since 
\begin{align}\label{eq:ydoublesing}
\int_0^{y_1} \rmd y\, y^{p-\frac{1}{2}} (y_1-y)^{-\frac{1}{2}}
 = y_1^{p}
\int_0^{1} \rmd t\, t^{p-\frac{1}{2}} (1-t)^{-\frac{1}{2}},
\end{align}
both of the remaining 
integrals are uniformly bounded in $x$, independently of the actual
values of $y_1$ and $y_2$.  The rough estimate proves that $\sup_x J(x)$
is uniformly bounded for all $\vep$, but it also proves
that if $x\in [0,\vep)$ or if $x\in (2\pi-\vep,2\pi)$,
then there is a constant $c'$ such that $J(x)\le c'\vep^p$.

Let us then consider the remaining case when $\vep$ is small (say $\vep<1$)
and $x\in [\vep,2\pi-\vep]$. Then $\sin\frac{x}{2}\ge \sin\frac{\vep}{2}$,
and thus Lemma \ref{th:mainFpmest} shows that every $(x,y)\in I^2$ 
for which
$\vep \le y\le y_1(x)-\vep$ or $y_2(x)+\vep\le y\le 2\pi-\vep$, belongs to
$X_\vep$. Therefore, such $y$ do not contribute to $J(x)$, and we can estimate
\begin{align}
& J(x) \le C^{-1/2} \left(\sin\frac{x}{2}\right)^{p} \Bigl(
\int_0^{\min(\vep,y_1)} \rmd y\, f_1(y)
+ \int_{\max(0,y_1-\vep)}^{y_1} \rmd y\, f_1(y)
\nonumber \\ & \qquad 
+ \int_{y_2}^{\min(2\pi,y_2+\vep)} \rmd y\, f_2(y)
+ \int_{\max(2\pi-\vep,y_2)}^{2\pi} \rmd y\, f_2(y)
\Bigr) .
\end{align}
Each of the four integrals can be estimated similarly to
(\ref{eq:ydoublesing}), which shows that they are  
bounded by  $c''\vep^p$ for some constant $c''$.  Therefore,
we have proven that also in this case $J(x)\le c \vep^p$ for some $c$.
This completes the proof of the Proposition.
\end{proof}

\begin{proofof}{Proposition \ref{th:operprop}} 
By Propositions \ref{th:K2compact} and \ref{th:K1compact}, the operators
$\omega K_1\omega$ and $\omega K_2\omega$ are compact, and thus also bounded. 
By Lemma \ref{th:Wtest}, $W$ is a bounded function, and thus
$\tilde{L}$ is a sum of a bounded multiplication operator and a compact
integral operator.  Thus $\tilde{L}$ is bounded,
and  then the argument in the beginning of the section,
based on Proposition \ref{th:Lbetaprop},
implies that it is positive.
On the other hand, Lemma \ref{th:Wtest} also implies that
$0\le V(x)^{-1/2} = \omega(x) W(x)^{-1/2} \le c_2 \sin(x/2)^{1/6}$, and we can
apply 
Propositions \ref{th:K2compact} and \ref{th:K1compact} also to the
definition of $B$, equation (\ref{eq:defkerB}).  This proves that
$B$ is a compact, self-adjoint operator on $L^2(I)$.
$B$ and $\tilde{L}$ also commute with $P$, by the symmetry properties of
$F_\pm$ stated in the beginning of the proof of Lemma \ref{th:Wtest}.

Thus we only need to prove the last claim in the Proposition.
Applying the definitions, we find
$\tilde{L}=W^{1/2} (1-B) W^{1/2}$.  For any $\psi\in L^2$, for which
$W^{-1/2} \psi \in L^2(I)$, we can find a sequence $f_n$ of
Lipschitz continuous functions,
such that $f_n \to  W^{-1/2}\psi$ in $L^2$.
Thus by the boundedness of $\tilde{L}$ and Proposition \ref{th:Lbetaprop},
then
\begin{align}
& \braket{\psi}{(1-B) \psi}
= \lim_n \braket{f_n}{\smash{\tilde{L}} f_n}
= \lim_n \int_{I^2} \rmd x \rmd z \frac{1}{2 \sqrt{F_+(x,z)}}
 \nonumber \\ & \qquad \times
  \left|g_n(x)+g_n(h(x,z))-g_n(z)-g_n(x-z+h(x,z))\right|^2,
\end{align}
where $g_n= \omega f_n \to V^{-1/2} \psi$. The function defined by the
integral on 
the right hand side is $L^2$-continuous in $f_n$. To see this, let us 
inspect the
difference of two such integrals, which can be bounded by a sum of finitely
many terms of the type
$\int\! \rmd x\rmd z (4 F_+)^{-1/2} |G(X)|^2$, where $X$
denotes any one of the functions $x$, $z$, $h(x,z)$, or $x-z+h(x,z)$,
and $G$ is in $L^2$.
The first two choices of $X$ lead to integrals which clearly can be
bounded by $\int \rmd x V(x) |G(x)|^2$.  However, 
so do the last two choices, as can be seen by employing
the symmetry $h(z,x)=x-z+h(x,z)$ and Lemma \ref{th:lemmaK2toK1}. 
Using then the fact that any relevant $G$ is of the form $G=\omega F$,
$F\in L^2$, we have here $\int \rmd x V(x) |G(x)|^2
= \int \rmd x W(x) |F(x)|^2\le \norm{W}_\infty \norm{F}^2$.  This suffices to
prove the continuity, and thus for the above class of $\psi$,
\begin{align}\label{eq:1minusBeq}
& \braket{\psi}{(1-B) \psi} 
= \int_{I^2} \rmd x \rmd z \frac{1}{2 \sqrt{F_+(x,z)}}
 \nonumber \\ & \qquad \times
  \left|g(x)+g(h(x,z))-g(z)-g(x-z+h(x,z))\right|^2,
\end{align}
with $g= V^{-1/2} \psi$. Then the previous argument can also be applied to show
that the right hand side is $L^2$-continuous in $\psi$, which proves that
(\ref{eq:1minusBeq}) holds for all $\psi\in L^2$.  Therefore,
$1-B\ge 0$, and $B\psi =\psi$ if and only if the integral on
the right hand side of (\ref{eq:1minusBeq}) vanishes for $g=V^{-1/2} \psi$.
Since then the integrand must be zero almost everywhere, this is possible if
and only if the periodic extension of $g$ is a collisional invariant.
\end{proofof}

\section{Collisional invariants (proof of Theorem \ref{th:collinv})}
\label{sec:provecolli}

It is clear that every $\psi(x)=c_1+c_2 \omega(x)$ is a locally integrable
collisional invariant.  Thus to prove the Theorem, it will be enough to
consider any $\psi$, which is a locally integrable collisional invariant,
and to show that it is almost everywhere equal to a function of the above
form.  Let us assume $\psi$ is such a function.  Then,
as $\Omega(x,h(x,z),z)=0$ for all $x,z$, 
we have for almost every $x,z\in \R$
\begin{align}\label{eq:collinveq}
\psi(x)+\psi(h(x,z))-\psi(z)-\psi(x-z+h(x,z)) =0.
\end{align}
In addition, since $h(2 \pi-x,2\pi-z)=-h(x,z)$ for $x,z\in I$, then also
$P\psi$ satisfies (\ref{eq:collinveq}) almost everywhere.  
Therefore, both of $(\psi\pm P\psi)/2$ have this property, and thus
it is sufficient to prove the result assuming that $\psi$ is either symmetric
of antisymmetric under $P$.

Let us begin by showing that then there
is $f:\R\to\C$ which is periodic and twice
continuously differentiable apart possibly from points in $2\pi\Z$, and
for which $\psi=f$ almost
everywhere.  We will do this  by integrating (\ref{eq:collinveq}) over $x$. 
However, the integration region has to be chosen with some care, in order to
guarantee that the result is finite.  With a certain
$0<\vep_0<\frac{\pi}{4}$ to be fixed later, we consider 
an arbitrary $0<\vep\le \vep_0$.
We define for all $\vep\le z\le \pi + \vep$
\begin{align}\label{eq:deff1}
f_{1,\vep}(z) = \frac{1}{\vep} \int_{2\pi-\vep}^{2\pi} \!\rmd x\, 
\left[\psi(x)+\psi(h(x,z))-\psi(x-z+h(x,z))\right].
\end{align}
A comparison with (\ref{eq:collinveq}) reveals that then 
$f_{1,\vep}(z)=\psi(z)$ almost everywhere.
On the whole integration region $x>z$ and thus 
\begin{align}\label{eq:hzlessx}
  h(x,z) = \frac{z-x}{2} + \Phi(x,z) \qand
  x-z+h(x,z) = \frac{x-z}{2} + \Phi(x,z) 
\end{align}
with
\begin{align}
\Phi(x,z) =   2 \arcsin \! \left( \tan \frac{x-z}{4} \cos \frac{x+z}{4}
\right) . 
\end{align}

We claim that there are $\vep_0,C>0$
such that for all $x,z,\vep$ as above
\begin{align}\label{eq:mainderivbound}
 2 \partial_x \Phi(x,z) \le -1-C.
\end{align}
Together with (\ref{eq:hzlessx}) 
this implies that the last two mappings in the arguments of $\psi$ 
in (\ref{eq:deff1}) are strictly
decreasing in $x$.  In particular, when $x\to 2\pi$, we have also
$h(x,z)\searrow z-2 \pi$ and $x-z+h(x,z)\searrow 0$.  Thus
a change of variables and denoting $\zeta_0=h(2 \pi -\vep,z)$ yields
\begin{align}\label{eq:deff1vep}
& f_{1,\vep}(z) = \frac{1}{\vep} \int_{2\pi-\vep}^{2\pi} \!\rmd x\, \psi(x)
+ \frac{1}{\vep} \int_{z-2\pi}^{\zeta_0} \!\rmd \zeta\, \psi(\zeta)
 \frac{2}{1-2\partial_x \Phi(x_1(\zeta,z),z)}
\nonumber \\ & \quad 
- \frac{1}{\vep} \int_{0}^{2\pi-\vep-z+\zeta_0} \!\rmd \zeta\, \psi(\zeta)
 \frac{2}{-1-2 \partial_x \Phi(x_2(\zeta,z),z)} .
\end{align}
Since both of the factors multiplying $\psi(\zeta)$ are
continuous in $z$ and uniformly bounded, we can conclude using the 
dominated convergence theorem 
that $f_{1,\vep}$ is continuous.  

As mentioned before, 
$\psi(z)=f_{1,\vep}(z)$ for almost every $\vep\le z\le \pi+\vep$.
Since $P\psi = \sigma \psi$, with $\sigma \in\set{\pm 1}$,
we can define $f_\vep(z) = f_{1,\vep}(z)$ for $\vep\le z\le \pi+\vep$,
and $f_\vep(z) = \sigma f_{1,\vep}(2\pi-z)$ for $\pi-\vep\le z\le 2\pi-\vep$.
In the common domain near $z=\pi$ both functions have to be equal
\defem{everywhere}, as they are continuous and coincide with $\psi$ almost
everywhere.  In particular, $f_\vep(z)$ is also everywhere continuous and
equal to $\psi$ almost everywhere. Since $\vep$ was arbitrary,
we can then extend the definition to cover the whole of 
$(0,2 \pi)$, by choosing 
$f(z)=f_\vep(z)$ for any $\vep<z,2\pi-z$.  
Again, by continuity, any two functions 
$f_\vep$ and $f_{\vep'}$ must agree on the intersection of their domains of
definition, so $f$ is a continuous function on $(0,2 \pi)$,
which we extend periodically to $\R$.  Then
$\psi(z)=f(z)$ a.e.\ $z\in \R$.  

Using the continuity of $h$, this implies that
\begin{align}\label{eq:fcollinveq}
f(x)+f(h(x,z))-f(z)-f(x-z+h(x,z)) =0,
\end{align}
for all $x,z\in I$ 
for which all arguments are non-zero, i.e., whenever $x\ne 0$, 
$z\ne 0$ and $x\ne z$.  In particular, then (\ref{eq:deff1vep}) holds
for all $\vep \le z \le \pi +\vep$ after
both $\psi$ and $f_{1,\vep}$ are replaced by $f$.
However, then the right hand side of (\ref{eq:deff1vep}) is continuously
differentiable, and we can conclude that $f$ is continuously
differentiable on $(0,2 \pi)$.  This argument can then be iterated once
more to conclude that $f$ must be twice continuously differentiable on 
$(0,2 \pi)$ (this way even smoothness could be proved, but we will not
need this property here). 

We next prove that we can choose $f(0)$ so that $f$ is continuous and
$f'(x)$ has a limit for both $x\searrow 0$ and for $x\nearrow 2\pi$.
Since $h(x,2 \pi-x) = \pi - x$ for all $x\in I$, we have for all
$x\in (0,\pi)$,
\begin{align}\label{eq:pix}
f(x)-f(2 \pi-x) +f(\pi-x) - f(\pi+x) =0.
\end{align}
If $f$ is antisymmetric, $f(\pi)=0$ and
(\ref{eq:pix}) implies that for all $x\in (0,\pi)$, $f(x) = f(\pi+x)$.
Therefore, in this case $f$ is continuously
differentiable at $x=0$, after we define $f(2\pi n)=0$, $n\in \Z$.

Assume then that $f$ is symmetric which implies $f'(z)=-f'(2\pi -z)$.
Let us consider values $0<z<x<2\pi$, when (\ref{eq:hzlessx}) holds.
Differentiating (\ref{eq:fcollinveq}) with respect to $x$ and $z$ yields
\begin{align}
  f'(x) + \partial_x h f'(h)- (1+\partial_x h) f'(x-z+h) & =0,  \\
  -f'(z) + \partial_z h f'(h)- (-1+\partial_z h) f'(x-z+h)& =0 .
\end{align}
We multiply the second equality by $(1+\partial_x h)$, and then use the first
one to eliminate $f'(x-z+h)$.  This proves that
\begin{align}\label{eq:fp2}
  (1-\partial_z h) f'(x) - (1+\partial_x h) f'(z) 
 + (\partial_x h +\partial_z h) f'(h) =0 .
\end{align}
We divide the equality by $1-\partial_z h$ and consider taking the limit
$x\to 2\pi$ for a fixed $z$.   
Then $h\to z-2\pi$, and the partial derivatives converge as (see
(\ref{eq:DG}) 
to obtain explicit formulae from which these can be checked)
\begin{align}\label{eq:Dhxlim}
\partial_x h(x,z) \to -(1+t^2) \qand \partial_z h(x,z) \to 1,
\end{align}
where $t=\tan \frac{2\pi -z}{4}$.  Since $1-\partial_z h\to 0$,
we need to compute the limit more carefully.  Let us fix for definiteness,
$z=\pi$, when a straightforward computation shows that 
$\partial_z \partial_x h(x,z) \to 1$, and thus by L'Hospital's rule,
\begin{align}
 \lim_{x\nearrow 2\pi}\frac{f'(z)-f'(h(x,z))}{1-\partial_z h(x,z)}
 =  \lim_{x}
 \frac{f''(h(x,z))\partial_x h(x,z)}{\partial_x\partial_z h(x,z)}
 =  -2 f''(\pi).
\end{align}
We can then use this result in (\ref{eq:fp2}) to prove that
the limit of $f'(x)$ exists when $x\nearrow 2\pi$, and thus by symmetry the
same is true about the limit when $x\searrow 0$.
Since this implies that $f'$ is bounded on $[0,2\pi]$, we can also conclude
that the limit $c=\lim_{x\to 0} f(x)$ exists, and we can make $f$
continuous by defining $f(2\pi n)=c$ for all $n\in\Z$.

We can thus assume that $f$ is continuous and periodic
on $\R$, continuously differentiable
on $(0,2\pi)$, and 
that $a=\lim_{x\searrow 0} f'(x)$ and $b=\lim_{x\nearrow 2\pi} f'(x)$
exist.  We also have $b=-a$, if $f$ is symmetric, and $b=a$, if $f$ is
antisymmetric. Let us now consider any $0<z\le \pi$, and $x=2\pi-\vep$
for $0<\vep<2\pi-z$.
As proven earlier, in the limit $\vep\searrow 0$, $x-z+h\searrow 0$,
and we get from (\ref{eq:fcollinveq})
that when $\vep\to 0$,
\begin{align}
& \frac{f(2\pi+h)-f(z)}{2\pi-x} = \frac{f(x+h-z)-f(0)+f(2\pi)-f(x)}{2\pi-x}
\nonumber \\ & \quad 
 \to -(1+\partial_x h) a +b.
\end{align}
The left hand side converges to $- f'(z)\partial_x h$,
and, since by (\ref{eq:Dhxlim}) $\partial_x h(2\pi,z)=-(1+t^2)<0$, we have
proven that
\begin{align}
 f'(z) = a - \frac{1}{1+t^2} (a-b) ,
\end{align}
where $\frac{1}{1+t^2} = \cos^2 \frac{2\pi -z}{4} = \frac{1}{2}\left(
1 - \cos\frac{z}{2}\right)=\frac{1}{2}-\omega'(z)$.  Therefore, for 
$0<z\le \pi$ we need to have
\begin{align}
 f'(z) = \frac{a+b}{2} + (a-b) \omega'(z).
\end{align}

If $f$ is antisymmetric, then $f'(z)=a$
for $0<z\le \pi$. Since then also $f'(2\pi-z)=a$, we must have
$f(z)=c+a z$ for $0<z< 2\pi$.  However, as also $f(0)=0=f(2\pi)$,
we need to have $c=0=a$, and thus the only antisymmetric 
solution is the trivial solution $f=0$. 
If $f$ is symmetric, $b=-a$, and $f'(2\pi-z)=-f'(z)$.  Thus then
$f'(z) = 2 a \omega'(z)$, for $0<z<2\pi$, and there is $c\in\C$ such that
$f(z)=c+2 a \omega(z)$ for all $z\in \R$.  Therefore, $f$ is in both cases a
trivial  collisional invariant, and since $\psi=f$ almost everywhere, 
we have arrived at the conclusion made in the Theorem.

We still need to prove (\ref{eq:mainderivbound}).  
Using the shorthand notations $t=\tan \frac{x-z}{4}$,
$c=\cos \frac{x+z}{4}$, we can write
\begin{align}\label{eq:DG}
& 2 \partial_x \Phi(x,z) +1 = \frac{c(1+t^2)-t\sqrt{1-c^2}}{\sqrt{1-c^2 t^2}} +1
\nonumber \\ & \quad 
=  \frac{1+t^2}{\sqrt{1-c^2 t^2}} \left( c + 
\frac{1-t^2}{1+t^2}
\frac{1}{t\sqrt{1-c^2}+\sqrt{1-c^2 t^2}} \right)  .
\end{align}
For $\vep,x,z$ as above, i.e., for
$0<\vep<\frac{\pi}{4}$, $\vep\le z\le \pi + \vep$, and $2\pi-\vep\le x \le
2\pi$, 
\begin{align}
\frac{\pi}{4}-\frac{\vep}{2} \le \frac{x-z}{4} \le 
\frac{\pi}{2}-\frac{\vep}{4}
 \qand
\frac{\pi}{2}+\frac{z-\vep}{4} \le \frac{x+z}{4} \le
\frac{3 \pi}{4} +\frac{\vep}{4}.
\end{align}
Thus $t \ge \tan \left(\frac{\pi}{4}-\frac{\vep}{2}\right) > 0$, and
with $c'= \cos \frac{\pi -\vep}{4}$
\begin{align}
& -1<-c' \le  c \le
\cos\left(\frac{\pi}{2}+\frac{z-\vep}{4}\right) 
= -\sin \frac{z-\vep}{4} \le  0 .
\end{align}
If also $z\le \frac{\pi}{2}-\vep$, then $x-z\ge \frac{3 \pi}{2}$ and thus
$t \ge \tan(3 \pi/8)>2$.  In this case, we can estimate the first term
using $c\le 0$, which yields
\begin{align}
& 2 \partial_x \Phi(x,z) +1 \le -\frac{t^2-1}{t+1} = -(t-1) \le -1.
\end{align}
Otherwise, $z-\vep>\frac{\pi}{2}-2\vep \ge \frac{\pi}{4}$.
If $t\ge 1$, we find 
\begin{align}
 2 \partial_x \Phi(x,z) +1 \le  c \le -\sin \frac{z-\vep}{4}\le
-\sin \frac{\pi}{16}<0. 
\end{align}
On the other hand, if $t<1$, then by
\begin{align}
 \frac{1-t^2}{1+t^2} = \cos \frac{x-z}{2} \le 
 \cos\left(\frac{\pi}{2}-\vep\right) = \sin \vep,
\end{align}
we have
\begin{align}
& 2 \partial_x \Phi(x,z) +1 \le \frac{1+t^2}{\sqrt{1-c^2 t^2}}
\Bigl( c +  \frac{1-t^2}{1+t^2}\frac{1}{\sqrt{1-(c')^2}} \Bigr)
\nonumber \\ & \quad 
 \le  \frac{1+t^2}{\sqrt{1-c^2 t^2}} \Bigl(
 -\sin \frac{\pi}{16}+\frac{\sin \vep}{\sin \frac{\pi-\vep}{4}}  \Bigr).
\end{align}
Since the term in the parenthesis approaches $-\sin \frac{\pi}{16}<0$,
when $\vep\to 0$, there is $\vep_0>0$ such that for all $0<\vep\le \vep_0$
the right hand side is less than $-\frac{1}{2}\sin \frac{\pi}{16}$.
Thus we can conclude that for all such $\vep$ Equation
(\ref{eq:mainderivbound}) 
holds at least with $C=\frac{1}{2}\sin \frac{\pi}{16}$.
This completes the proof of the Theorem.

\section{Resolvent expansion  (proof of Theorem \ref{th:mainres})}
\label{sec:resolv}

Let us begin with the following corollary of the results proven in the
previous sections. 
\begin{corollary}\label{th:Bonecorr}
The eigenspace of $B$ with
eigenvalue $1$ is two-dimensional, and it is spanned by 
$V(x)^{1/2}$ and $\omega(x) V(x)^{1/2}$.  Every
$\psi \in L^2(I)$, such that $P\psi=-\psi$,
is orthogonal to this eigenspace.
\end{corollary}
\begin{proof}
Suppose $\psi$ belongs to the above eigenspace of $B$.
By Proposition \ref{th:operprop}, then 
$\tilde{\psi} = V^{-1/2} \psi= \omega W^{-1/2} \psi$ is a collisional invariant.
Thus by Theorem \ref{th:collinv}, there are $c_1,c_2\in \C$ such that 
$\omega  W^{-1/2} \psi=c_1 \omega +c_2$.  By Lemma \ref{th:Wtest},
both of the vectors $W^{1/2} = \omega V^{1/2}$ and 
$\omega^{-1}  W^{1/2}=V^{1/2}$ belong to $L^2$ and are symmetric under $P$.   
This proves the results stated in the Corollary.
\end{proof}

To estimate error terms, we will rely on the following estimates:
\begin{lemma}\label{th:basicint}
There is $C>0$ such that for any $0<\lambda<1$, 
\begin{align}\label{eq:basicint}
  \int_0^{2\pi}\rmd x\, \frac{\omega(x)}{W(x)+\lambda}
    \left(\sin \frac{x}{2}\right)^{-\frac{1}{2}} \le C \lambda^{-\frac{1}{10}}.
\end{align}
\end{lemma}
\begin{proof}
Let $0<\lambda<1$ be arbitrary.  By symmetry, the integrals over 
$[0,\pi]$ and $[\pi,2\pi]$ are equal.  On the other hand, 
by Lemma \ref{th:Wtest}, we have
\begin{align}\label{eq:basicint2}
  \int_0^{\pi}\rmd x\, \frac{\omega(x)}{W(x)+\lambda}
    \left(\sin \frac{x}{2}\right)^{-\frac{1}{2}} \le
\frac{1}{1+c_1}  \int_0^{\pi}\rmd x\, \frac{s_x^{1/2}}{s_x^{5/3}+\lambda}.
\end{align}
The integral over 
$x\in [\frac{1}{2}\pi, \pi]$ is clearly bounded uniformly in $\lambda$.
To estimate the integral over $[0,\frac{1}{2}\pi]$, we change the integration
variable to $s=\lambda^{-3/5} s_x$, which shows that
\begin{align}
  \int_0^{\pi/2}\rmd x\, \frac{s_x^{1/2}}{s_x^{5/3}+\lambda}
\le  2\sqrt{2} \lambda^{\frac{3}{5}+ \frac{3}{10} -1}
 \int_0^{\infty}\rmd s\, \frac{s^{1/2}}{s^{5/3}+1} \le c \lambda^{-\frac{1}{10}},
\end{align}
since the integral over $s$ is finite. Thus (\ref{eq:basicint})
holds for some finite $C$. 
\end{proof}

\begin{lemma}\label{th:supvarphi}
For $0<\lambda<1$, let
\begin{align}
  \varphi_\lambda = B \frac{W^{\frac{1}{2}}}{W+\lambda}\omega' .
\end{align}
For any $0<\vep<\frac{3}{10}$ there is a constant $c_\vep>0$ such that 
for all $x\in I$ and $\lambda$,
\begin{align}\label{eq:supvarphi}
 |\varphi_\lambda(x)| \le c_\vep \lambda^{-\frac{1}{10}-\vep} 
 \left(V(x) \sin \frac{x}{2}\right)^{-\frac{1}{2}}.
\end{align}
\end{lemma}
\begin{proof}
Let $0<\vep<\frac{3}{10}$ be arbitrary.
Applying the definitions of $B$ and
$W$, as well as the bound $|\omega' |\le \frac{1}{2}$, we find
\begin{align}
&  |\varphi_\lambda(x)| \le
 \frac{1}{2} V(x)^{-1/2} \int_0^{2\pi} \rmd y\, 
 \left(|K_1(x,y)|+ 2 |K_2(x,y)| \right)
    \frac{\omega(y)}{W(y)+\lambda}.
\end{align}
By Lemma \ref{th:basicint} and Eq.\ (\ref{eq:K2bound}), 
the second term in the sum satisfies
\begin{align}
\int_0^{2\pi} \rmd y\, 2 |K_2(x,y)|\frac{\omega(y)}{W(y)+\lambda}
\le C \left(\sin\frac{x}{2}\right)^{-\frac{1}{2}}
 \lambda^{-\frac{1}{10}},
\end{align}
and thus leads to a bound of the desired form.

To the first term we apply
Lemma \ref{th:mainFpmest}, which shows that
\begin{align}
& \int_0^{2\pi} \rmd y\,  |K_1(x,y)|
    \frac{\omega(y)}{W(y)+\lambda}
\le C
 \left|\sin\frac{x}{2}\right|^{-\frac{1}{2}} 
\nonumber \\ & \qquad \times
\Bigl[\int_0^{y_1(x)}\rmd y
\frac{1}{\sqrt{y_1(x)-y}}  
  \frac{\omega(y)}{W(y)+\lambda}
 + \int_{y_2(x)}^{2\pi}\rmd y \frac{1}{\sqrt{y-y_2(x)}} 
  \frac{\omega(y)}{W(y)+\lambda} \Bigr] .
\end{align}
Changing the integration variable to $y'=2\pi-y$ in the second integral
reveals that it is equal to the first integral, if $y_1(x)$
is replaced by $2\pi-y_2(x)$.  
Thus it is sufficient to inspect the first
integral.  We estimate it using Lemma \ref{th:Wtest} and H\"{o}lder's
inequality for $p'=\frac{6}{3+10\vep}<2$, $q'= \frac{6}{3-10\vep} >2$.
This shows that for any $y_1\in I$,
\begin{align}\label{eq:sqrteW}
& \int_0^{y_1}\rmd y \frac{1}{\sqrt{y_1-y}} 
  \frac{\omega(y)}{W(y)+\lambda}
\nonumber \\ & \quad
 \le \frac{1}{1+c_1}
\left[\int_0^{y_1}\rmd y (y_1-y)^{-\frac{p'}{2}}\right]^{\frac{1}{p'}}
\left[\int_0^{2\pi}\rmd y \left(\frac{\omega(y)}{
      \omega(y)^{5/3}+\lambda}\right)^{q'}
\right]^{\frac{1}{q'}}.
\end{align}
The first factor is an $\vep$-dependent, finite 
constant, and the second factor can be 
estimated as in the proof of Lemma \ref{th:basicint}: first we use symmetry to
reduce the estimate to $[0,\pi]$, then $\omega(y)>0$ on
$[\pi/2,\pi]$ to bound the integral on this interval by a constant, and finally
on the interval $[0,\pi/2]$ we change the integration variable to 
$s= \lambda^{-3/5} \sin\frac{y}{2}$.  This shows that there is a constant $c$
such that
\begin{align}
\int_0^{2\pi}\rmd y \left(\frac{\omega(y)}{\omega(y)^{5/3}+\lambda}\right)^{q'}
\le c \lambda^{\frac{3}{5}+ q'(\frac{3}{5} -1)}
 \int_0^{\infty}\rmd s\, \left(\frac{s}{s^{5/3}+1}\right)^{q'} \le c' 
\lambda^{-\frac{2}{5}q'+ \frac{3}{5}} ,
\end{align}
where the integral over $s$ is finite, since $q'>2>\frac{3}{2}$.
Therefore, as $-\frac{2}{5}+ \frac{3}{5q'}=-\frac{1}{10}- \vep$,
(\ref{eq:sqrteW}) implies that 
(\ref{eq:supvarphi}) holds also for the first term in
the sum.
\end{proof}

This has the following immediate corollary:
\begin{corollary}\label{th:varphiprop}
For any $0<\vep<\frac{3}{5}$ there is a constant $c_\vep>0$ such that  
$\norm{\varphi_\lambda}^2\le c_\vep \lambda^{-\frac{1}{5}-\vep}$
for all $0<\lambda<1$. In addition, 
$P\varphi_\lambda= -\varphi_\lambda$.  
\end{corollary}
\begin{proof}
Let $\vep'=\vep/2$.  Then by Lemma \ref{th:supvarphi},
\begin{align}
\norm{\varphi_\lambda}^2 = \int_0^{2\pi} \rmd x \, |\varphi_\lambda(x)|^2
\le  c'_{\vep'} \lambda^{-\frac{1}{5}-2 \vep'} \int_0^{2\pi} \rmd x \,
 \left(V(x) \sin \frac{x}{2}\right)^{-1}.
\end{align}
By Lemma \ref{th:Wtest}, there is $C$ such that 
$V(x) \sin (x/2)\ge C (\sin (x/2))^{2/3}$, and thus the remaining integral over
$x$ is finite.  This implies that there is $c_\vep$ such that 
$\norm{\varphi_\lambda}^2\le c_\vep \lambda^{-\frac{1}{5}-\vep}$.  
Since both $B$ and $W$ commute with
$P$, and $\omega'$ is antisymmetric, it follows that $\varphi_\lambda$ is
antisymmetric. 
\end{proof}

Armed with the above results, we can prove the main theorem.
We claim that for any $0<\vep<\frac{\alpha}{2}$,
\begin{align}\label{eq:Rzexpans}
& R(\lambda ) = \bigbraket{\omega' }{\smash[t]{\frac{1}{\lambda +W}} \omega' }
 + \order{\lambda^{-\frac{\alpha}{2}-\vep}}.
\end{align}
This implies that only the first term of a resolvent expansion
needs to be considered for the 
limit (\ref{eq:Rlimit}).  However, then
\begin{align}
\bigbraket{\omega' }{\smash[t]{\frac{\lambda^{\alpha}}{\lambda +W}} \omega' }
= \frac{1}{4}
\int_{0}^{2\pi}\!\rmd x\,
\frac{\lambda^{\frac{2}{5}}}{\lambda +W(x)} \cos^2\frac{x}{2},
\end{align}
and we  can use the symmetry of the integrand to reduce the integration region
to $[0,\pi]$, while gaining a factor of $2$.
We then change variables to $s= \lambda^{-3/5} \sin\frac{x}{2}$.
By Lemma \ref{th:Wtest}, the remaining integrand is dominated by 
$\frac{1}{1+c_1 s^{5/3}}$ which is integrable on $(0,\infty)$.  Thus
\begin{align}
\lim_{\lambda\to 0^+}
\bigbraket{\omega' }{\smash[t]{\frac{\lambda^{\alpha}}{\lambda +W}} \omega' } =
c_0 =  \int_{0}^\infty\! \rmd s\, \frac{1}{1+w_0 s^{\frac{5}{3}}}.
\end{align}
Clearly, $0<c_0<\infty$, as claimed in the Theorem.

To estimate the ``error term'' in (\ref{eq:Rzexpans}) assume $0<\lambda<1$ is
given.  Let $A=W^{\frac{1}{2}} B W^{\frac{1}{2}}$, when $A$ is a bounded
operator and $\tilde{L}= W-A$.
We use the resolvent expansion of $\tilde{L}$ up to the second order,
\begin{align}\label{eq:resolv2}
& \frac{1}{\lambda +\tilde{L}} = \frac{1}{\lambda +W} + 
\frac{1}{\lambda +W} A \frac{1}{\lambda +W}
  + \frac{1}{\lambda +W} A \frac{1}{\lambda +\tilde{L}} A \frac{1}{\lambda +W}
\end{align}
where, since $W\ge 0$, $(\lambda +W)^{-1}$ is a bounded, positive operator.
Therefore, denoting $\phi_\lambda = A \frac{1}{\lambda +W}\omega' $,
\begin{align}\label{eq:Rzexpans2}
& R(\lambda ) = \bigbraket{\omega' }{\smash[t]{\frac{1}{\lambda +W}} \omega' }
 + \bigbraket{\omega' }{\smash[t]{\frac{1}{\lambda +W}}A
      \smash[t]{\frac{1}{\lambda +W}} \omega' }
 + \bigbraket{\phi_\lambda }{
   \smash[t]{\frac{1}{\lambda +\tilde{L}}} \phi_\lambda } .
\end{align}
Using Proposition \ref{th:operprop}, Corollary \ref{th:Bonecorr},
and Lemmas \ref{th:basicint}--\ref{th:varphiprop}, 
we can now prove (\ref{eq:Rzexpans}).
To estimate the first correction we use Lemmas \ref{th:basicint}
and \ref{th:supvarphi}:
\begin{align}
& \left| \bigbraket{\omega' }{\smash[t]{\frac{1}{\lambda +W}}A
      \smash[t]{\frac{1}{\lambda +W}} \omega' }\right|
= \left|\bigbraket{\smash[t]{\frac{W^{\frac{1}{2}}}{W+\lambda}}\omega' }{
    \varphi_\lambda}\right|
\nonumber \\ & \quad 
\le \int_0^{2\pi} \rmd x\, \frac{\omega(x)}{W(x)+\lambda} V(x)^{\frac{1}{2}}
|\varphi_\lambda(x)|
\le  c_\vep C \lambda^{-\frac{1}{5}-\vep} .
\end{align}
This proves that the first correction is of the
claimed order, and we only need to inspect the final term in
(\ref{eq:Rzexpans2}).  

Firstly, 
$\phi_\lambda = W^{\frac{1}{2}}\varphi_\lambda$, and thus
\begin{align}
& \bigbraket{\phi_\lambda }{\smash[t]{\frac{1}{\lambda +\tilde{L}}}
  \phi_\lambda } 
= \bigbraket{\varphi_\lambda }{\smash[t]{\frac{1}{1-B+\lambda  W^{-1}}}
  \varphi_\lambda }. 
\end{align}
Here, by Lemma \ref{th:Wtest}, we have the operator inequalities
\begin{align}
1-B+\lambda  W^{-1} \ge 1-B+\frac{\lambda }{c_2} \ge \frac{\lambda }{c_2}>0.
\end{align}
where $\frac{\lambda }{c_2}$ is proportional to the unit operator, and thus
commutes with $B$. Therefore,
\begin{align}
(1-B+\lambda  W^{-1})^{-1} \le (1-B+\frac{\lambda }{c_2})^{-1}
\end{align}
implying
\begin{align}\label{eq:Rerrorest}
&0 \le \bigbraket{\phi_\lambda }{\smash[t]{\frac{1}{\lambda +\tilde{L}}} 
  \phi_\lambda }
\le \bigbraket{\varphi_\lambda }{\smash[t]{\frac{1}{1+\lambda /c_2-B}} 
  \varphi_\lambda }. 
\end{align}
By Proposition \ref{th:operprop}, $B$ is compact, and its
spectrum consists of isolated eigenvalues (apart from zero).  Since also 
$B\le 1$,  we can order the eigenvalues so 
that $1=\lambda_1>\lambda_2>\ldots$.
In particular, then $\delta=1-\lambda_2>0$.  Since $\varphi_\lambda $ is
antisymmetric, it is orthogonal to the eigenspace of $B$ with eigenvalue
$1$  by Corollary \ref{th:Bonecorr}.  
Therefore, using the spectral decomposition of $B$, we find that
\begin{align}
&\bigbraket{\varphi_\lambda }{\smash[t]{\frac{1}{1+\lambda /c_2-B}}
  \varphi_\lambda } 
\le \frac{1}{\delta+\lambda /c_2} \norm{\varphi_\lambda }^2
\le \frac{c}{\delta} \lambda ^{-\frac{1}{5}-\vep},
\end{align}
where we used the estimate in Corollary \ref{th:varphiprop}.
Then we can conclude from  (\ref{eq:Rerrorest}) that
(\ref{eq:Rzexpans}) holds.  This completes the proof of Theorem
\ref{th:mainres}.

\appendix 

\section{Integral operators}
\label{sec:intkernels}

Given a positive measure $\mu$ on $X$, any function 
$K:X\times X\to \C$, which is measurable in $\mu\times \mu$,
can be used to define an operator $T$ in $L^2(\mu)$ by the formula
\begin{align}
  (Tf)(x) = \int_X \mu(\rmd y) K(x,y) f(y).
\end{align}
More precisely, we define $T$ as a possibly unbounded operator with the domain
\begin{align}
  D(T) = \defset{f\in L^2(\mu)}{\int \mu(\rmd x) \left(\int \mu(\rmd y)
      |K(x,y)| \,|f(y)| \right)^2 <\infty }.
\end{align}
$K$ is then called the integral kernel of the integral operator $T$.

We need here only the following convenient estimate for an operator 
norm of such integral operators.
\begin{proposition}\label{th:Anormcor}
Let $\mu$ be a positive measure on $X$, and assume that
$A:X\times X\to \C$ is measurable with respect to $\mu\times \mu$ and
satisfies $A(x,y)^*=A(y,x)$ for almost every $(x,y)\in X^2$.
Consider any measurable $\phi : X\to \C$, let 
$B(x,y) = \phi(x)^* A(x,y)\phi(y)$, and let $T$ denote 
the corresponding integral operator.  If there exists 
$\alpha\in \R$ such that
\begin{align}
  C_{\alpha} 
  = \esssup_x \left(|\phi(x)|^{2-\alpha} 
    \int_X \mu(\rmd y)\, |A(x,y)|\, |\phi(y)|^{\alpha}\right)   < \infty,
\end{align}
then $T$ is a bounded, self-adjoint operator on
$L^2(\mu)$, and  $\norm{T}\le C_{\alpha}$. 
\end{proposition}
\begin{proof}
Let $\alpha\in \R$ be given, and let us denote $\alpha'=2-\alpha$.  Then
$|\phi(x)|^2= |\phi(x)|^\alpha|\phi(x)|^{\alpha'}$. (We apply here the
usual convention used in connection with positive measures, that $0\cdot
\infty = 0$. 
Let also $0^0=1$).  
Thus for any $f\in L^2$, we have by the Schwarz inequality and Fubini's theorem
\begin{align}\label{eq:fAAf}
& \int_{X^3} (\mu\times \mu\times \mu)(\rmd (x,y,z))
|f(x)|\, |B(z,x)|\, |B(z,y)| \, |f(y)| 
\nonumber \\ & \quad 
 = \int_{X^3} (\mu\times \mu\times \mu)(\rmd (x,y,z))
\nonumber \\ & \qquad 
\times
|f(x)| \left(|A(z,x)|\, |A(z,y)|\, |\phi(x)|^{\alpha'} 
|\phi(z)|^{\alpha+\alpha'} |\phi(y)|^{\alpha} \right)^{\frac{1}{2}}
\nonumber \\ & \qquad 
\times  |f(y)|  \left(|A(z,x)|\, |A(z,y)|\, |\phi(y)|^{\alpha'} 
|\phi(z)|^{\alpha+\alpha'} |\phi(x)|^{\alpha} \right)^{\frac{1}{2}}
\nonumber \\ & \quad 
\le \int \mu(\rmd x) |f(x)|^2 \Bigl(
|\phi(x)|^{\alpha'} \int \mu(\rmd z) |A(x,z)|\,
|\phi(z)|^{\alpha} \,
\nonumber \\ & \qquad 
\times \Bigl[ |\phi(z)|^{\alpha'} \,
\int \mu(\rmd y) |A(z,y)|\, |\phi(y)|^{\alpha} \Bigr] \Bigr)
 \nonumber \\ & \quad 
 \le C_\alpha^2 \norm{f}^2 < \infty,
\end{align}
where we have used the symmetry of $A$.
Therefore, $D(T)=L^2$,
and since the left hand side of (\ref{eq:fAAf}) is an upper bound for 
$\norm{T f}^2$, we have proven that $\norm{T}\le C_\alpha$.
As $B(x,y)^*=B(y,x)$ almost everywhere,  $T$ is then also self-adjoint.
\end{proof}


\end{document}